\documentclass{article}

\usepackage{microtype}
\usepackage{graphicx}
\usepackage{subfigure}
\usepackage{booktabs} 

\usepackage{hyperref}

\usepackage[accepted]{icml2024}

\usepackage[utf8]{inputenc} 
\usepackage[T1]{fontenc}    
\usepackage{hyperref}       
\usepackage{url}            
\usepackage{booktabs}       
\usepackage{amsfonts}       
\usepackage{nicefrac}       
\usepackage{microtype}      
\usepackage{xcolor}         

\usepackage{microtype}
\usepackage{graphicx}
\usepackage{subfigure}
\usepackage{booktabs} 
\usepackage{diagbox}
\usepackage{hyperref}
\usepackage{caption}
\usepackage{afterpage}

\usepackage{ulem}

\usepackage{amsmath}
\usepackage{amssymb}
\usepackage{amsbsy}
\usepackage{mathtools}
\usepackage{amsthm}
\usepackage{bm}
\usepackage{adjustbox}
\usepackage{tabularx}
\usepackage{adjustbox}
\usepackage{bbding}
\usepackage{transparent}
\usepackage{multirow}
\usepackage[capitalize,noabbrev]{cleveref}

\usepackage{array}
\newcolumntype{?}{!{\vrule width 1.2pt}}
\makeatletter
\newcommand{\thickhline}{%
    \noalign {\ifnum 0=`}\fi \hrule height 1.3pt
    \futurelet \reserved@a \@xhline
}

\DeclareMathOperator*{\argmin}{argmin}

\theoremstyle{plain}

\theoremstyle{definition}

\theoremstyle{remark}

\usepackage[textsize=tiny]{todonotes}

\icmltitlerunning{Submission and Formatting Instructions for ICML 2024}

\begin{document}

\twocolumn[
\icmltitle{Auto-Linear Phenomenon in Subsurface Imaging}



\icmlsetsymbol{equal}{*}

\begin{icmlauthorlist}
\icmlauthor{Yinan Feng}{unccs}
\icmlauthor{Yinpeng Chen}{google}
\icmlauthor{Peng Jin}{Penn State}
\icmlauthor{Shihang Feng}{lanl}
\icmlauthor{Youzuo Lin}{uncds}
\end{icmlauthorlist}

\icmlaffiliation{unccs}{Department of Computer Science, The University of North Carolina at Chapel Hill,USA}
\icmlaffiliation{uncds}{School of Data Science and Society, The University of North Carolina at Chapel Hill,USA}
\icmlaffiliation{google}{Google Research, USA}
\icmlaffiliation{Penn State}{College of Information Sciences and Technology, The Pennsylvania State University, USA}
\icmlaffiliation{lanl}{Earth and Environmental Sciences Division, Los Alamos National Laboratory,USA}

\icmlcorrespondingauthor{Youzuo Lin}{yzlin@unc.edu}

\icmlkeywords{Machine Learning, ICML}

\vskip 0.3in
]



\printAffiliationsAndNotice{}  

\begin{abstract}


Subsurface imaging involves solving full waveform inversion (FWI) to predict geophysical properties from measurements. This problem can be reframed as an image-to-image translation, with the usual approach being to train an encoder-decoder network using paired data from two domains: geophysical property and measurement. A recent seminal work (InvLINT) demonstrates there is only a linear mapping between the latent spaces of the two domains, and the decoder requires paired data for training.

This paper extends this direction by demonstrating that only linear mapping necessitates paired data, while both the encoder and decoder can be learned from their respective domains through self-supervised learning. This unveils an intriguing phenomenon (named Auto-Linear) where the self-learned features of two separate domains are automatically linearly correlated. Compared with existing methods, our Auto-Linear has four advantages: (a) solving both forward and inverse modeling simultaneously, (b) applicable to different subsurface imaging tasks and achieving markedly better results than previous methods, (c)enhanced performance, especially in scenarios with limited paired data and in the presence of noisy data, and (d) strong generalization ability of the trained encoder and decoder.

\end{abstract}

\section{Introduction}
\label{Introduction}

\begin{figure}[ht]
\vskip 0.2in
\begin{center}
\centerline{\includegraphics[width=\columnwidth]{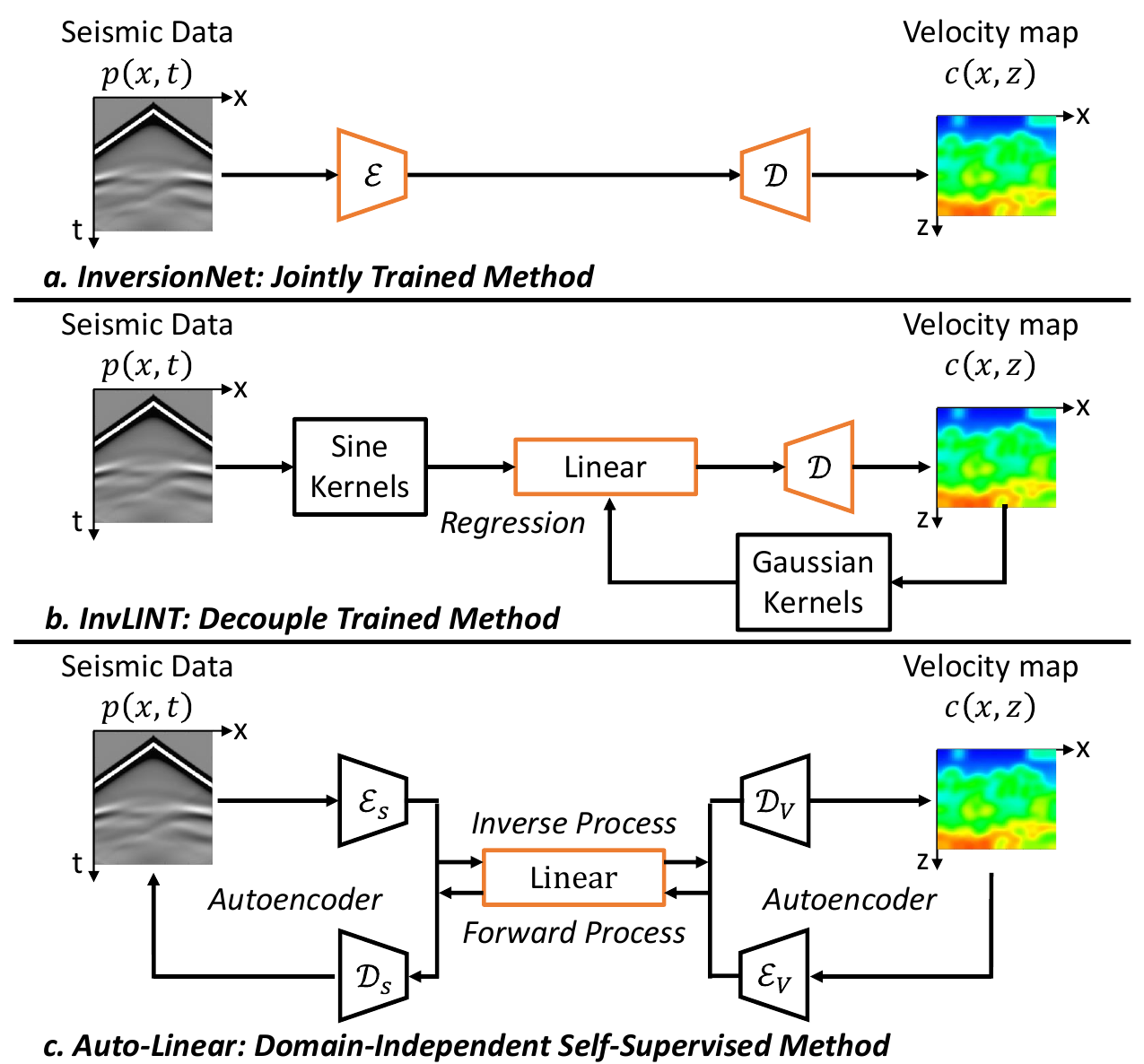}}
\caption{
\textbf{Overview} of the jointly trained encoder-decoder (top), InvLINT (middle), and Auto-Linear (bottom). The orange color indicates that components need to be trained with paired data. Auto-Linear decouples both the encoder and decoder and self-supervised trains them separately in their own domains. Linear converters are learned to connect the frozen, pre-trained encoders and decoders.
}
\label{teaser}
\end{center}
\vskip -0.4in
\end{figure}


Subsurface imaging is crucial for revealing subsurface layering and geophysical properties~(such as velocity and conductivity), supporting important applications such as energy exploration, carbon capture, and earthquake early warning systems. In this field, the full waveform inversion~(FWI) is a well-known method to infer subsurface velocity maps from the seismic data. Concurrently, the forward process involves computing the pressure wavefield using these velocity maps. Specifically, seismic data are obtained via seismic surveys which employ receivers to record reflected and refracted seismic waves generated by controlled sources. Each receiver records a 1D time series signal, and the signals recorded by all the receivers form the seismic data. They are mathematically connected by an acoustic wave equation as:
\begin{equation}
 \nabla^2p(x,z,t) -\frac{1}{c^2(x,z)} \frac{\partial^2}{\partial t^2}p(x,z,t) = s(x,z,t), \label{Acoustic}
\end{equation}
where $p(x,z,t)$ represents the seismic data and $c(x,z)$ is the velocity map. $s(x,z, t)$ is the source term. $x$ is the horizontal direction, $z$ is the depth, $t$ denotes time, and $\nabla^2$ is the Laplacian operator. In practice, seismic data is typically collected by surface sensors (i.e., $p(x, z=0, t)$, abbreviated as $p(x, t)$). 
This inversion problem is ``ill-posed,'' presenting challenges due to sensitivity to initial conditions and potential for multiple solutions. Furthermore, the substantial costs and logistical hurdles of acquiring real subsurface data often make it infeasible to gather extensive real-world datasets. Consequently, much of the current research primarily depends on full-physics simulations for experimental work, driven by the scarcity of publicly available real datasets.




Recent works \cite{wu2019inversionnet, zhang2019velocitygan, sun2021physics, jin2021unsupervised} consider FWI as an \textit{image-to-image translation problem} constrained by a wave equation, and leverage deep neural networks to achieve a significant performance boost. As shown in Figure~\ref{teaser}-a, they learn an encoder-decoder architecture to map seismic data to velocity. Note that the encoder and decoder are jointly trained from the supervision of paired seismic data and velocity maps. 


In \citet{feng2022intriguing}, the authors of InvLINT laid the foundation for exploring linear relationships in latent spaces by separating encoders and the decoder into two distinct domains and retaining a linear component for joint training. Shown in Figure~\ref{teaser}-b, their approach utilizes two predetermined integral transforms, with Sine and Gaussian kernels, to encode seismic data and velocity maps. While the encoders are decoupled, the decoder still relies on supervised training due to the absence of an explicit inversion for the Gaussian integral transform. This also restricts the decoder's ability to generalize among multiple datasets. Moreover, the kernel solution faces notable limitations: a) poor performance with datasets containing large variations and high-frequency components (e.g., OpenFWI~\cite{dengopenfwi}), b) a lack of noise resistance, and c) no clear rule for selecting the appropriate kernels for different situations. Thus, InvLINT may not apply to broader scenarios.



In this paper, we provide a more modular framework for data-driven subsurface imaging that can overcome all the limitations of InvLINT via self-supervised learning. This approach fully decouples the training of the encoder and decoder. Specifically, we first independently train two masked autoencoders (MAEs)~\cite{he2022masked} separately, one for seismic data and another for velocity maps, as depicted in Figure~\ref{teaser}-c. After self-pretraining, the encoder and decoder are frozen, and a linear converter is then trained to connect these components using paired seismic data and velocity maps. This efficient framework uniquely enables us to address both the inverse problem and forward modeling simultaneously, with minimal additional training required. Moreover, it exhibits considerable improvement in few-shot scenarios with paired data, presenting a substantial advancement over traditional image-to-image translation-based inversion networks like InversionNet~\cite{wu2019inversionnet}.


Moreover, we formula the above phenomenon that \textit{the \textbf{independently} self-supervised encoder and decoder of two different domains can be \textbf{automatically} integrated into an end-to-end model through a supervised \textbf{linear} mapping} as \textit{Auto-Linear Phenomenon}. The phenomenon offers an intriguing insight: an inherent and strong cross-domain correlation, with a simpler linear mapping, exists within the domain-independent self-consistent representations. This introduces a significant change in our perspective of the problem, moving away from the conventional, complex image-to-image translation task towards a more simplified, linear approach achieved through self-supervised learning. This modification not only streamlines the process but also enhances our comprehension of the auto-alignment of representation learned at each domain, offering a deeper insight into the interplay of data relationships across domains.

In experiments, Auto-Linear achieves solid performance on multiple datasets. Compared with joint training methods (e.g., InversionNet~\cite{wu2019inversionnet}), Auto-Linear has comparable results in the inverse problem, with its superiority in forward modeling and few-shot contexts with limited paired data and enhanced noise robustness. Moreover, Auto-Linear outperforms the previous decouple-trained method, InvLINT, and exhibits greater robustness to noisy data. Furthermore, it excels over both InvLINT and InversionNet in electromagnetic (EM) inversion, another subsurface imaging task. Our framework also enhances our understanding of relationships among multiple FWI datasets with distinct subsurface structures. We found that while these datasets can share both encoders and decoders, they require different linear mappings. In addition, we observe a correlation between the linear layer's singular values and the complexity of the dataset. Essentially, these suggest a consistent embedding method across datasets, with varying subsurface characteristics captured by cross-domain linear mappings. This leads to a piece-wise linearity between the two domains. 

\section{Related Works}
Recently, data-driven methods for FWI have been developed. They consider the FWI as an image-to-image problem and jointly train the encoder-decoder network to solve it. \citet{araya2018deep} use a fully connected network to invert velocity maps. \citet{wu2019inversionnet} adopted an encoder-decoder CNN to solve. \citet{zhang2019velocitygan} employ GAN and transfer learning to improve the generalization. In \citet{zeng2021inversionnet3d}, authors present an efficient and scalable encoder-decoder network for 3D FWI. \citet{feng2021multiscale} develop a multi-scale framework with two convolutional neural networks to reconstruct the low- and high-frequency components of velocity maps. A thorough review of deep learning for FWI can be found in \citet{Physics-2023-Lin}.

\citet{jin2021unsupervised} use the finite difference to approximate the forward modeling as a differentiable operator and integrate it and a deep neural network (DNN) in a loop to construct an unsupervised learning method. \citet{chen2021equivariant} proposed a self-supervised approach to solve the inverse problem from the perspective of image invariance. These purely self-supervised and unsupervised methods focus on how to solve problems without labels and still treat the network as a black box. Unlike them, our method uses self-supervised learning as a tool with the aim of simplifying the problem and decoupling the inverse process. We hope this can help the field better understand the problem and the relationship among different subsurface structures.

Recently, OpenFWI was released. It is the first open-source collection of large-scale multi-structural benchmark datasets for FWI~\cite{dengopenfwi}. It includes 12 datasets (11 2D datasets and one 3D dataset) synthesized from multiple sources. The datasets cover diverse domains in geophysics, such as interfaces, faults, and CO2 reservoirs, and feature a variety of subsurface structures, including flat and curved geologies. Along with the dataset, they also report performance benchmarks by using state-of-art data-driven methods and the physics-driven method. 


An alternative self-supervised approach for inverse problems involves pre-training a generative model on physical properties to capture their prior distribution. Subsequently, the model is adapted by integrating measurements and a physical model of the measurement process into the sampling process~\cite{wang2023prior, song2021solving}. The proposed Auto-Linear and generative model-based methods offer \textbf{complementary insights} into inverse problems, each revealing the problem from different perspectives. Auto-Linear leverages self-supervised learning to streamline connecting seismic data with subsurface velocity maps, focusing on understanding the latent space properties of two physical quantities connected via PDE and enriching our understanding of their relationships. Conversely, generative models focus on capturing the data's complex distributions to the entire space of possible solutions. They excel in integrating the forward modeling into sampling, and only relying on samples from prior distributions without necessitating paired samples.

A key insight of our method is the ability of self-learned representations, generated by an autoencoder in one domain, to retain essential information for accurate reconstruction in another domain via minimum transformation. The crucial aspect here is the \textbf{automatic} alignment of representations across domains, while the linear relationship emphasizes the simplicity of this relationship and its strong correlation. Thus, our work provides a new perspective on solving FWI problems. It presents a complementary, yet distinct, approach to existing generative model-based approaches.

Table~\ref{table:Generative} presents a side-by-side comparison between our approach and generative model-based approaches. First, Generative model-based approaches are trained in a purely self-supervised manner do not require paired data, while our approach necessitates paired data for training the linear converter. Second, unlike diffusion models that require multi-step denoising, our setup enables efficient single-step inference. Third, we target a conditional average with minimum error, focusing on a singular outcome based on the training data's distribution, contrasting with generative models-based methods that aim to explore multiple potential solutions through distribution-to-distribution mappings. Additionally,  generative model-based methods integrate forward modeling into the learning process and benefit from existing knowledge of physical processes; however, they struggle in situations where the forward model is unknown or lacks an explicit formulation. For instance, Kimberlina carbon sequestration problem~\cite{alumbaugh2021development}, while data, generated by Maxwell’s Equations, has been made available, the forward modeling remains proprietary and challenging to replicate. In such scenarios, our approach presents an effective alternative, offering a different way for inverse analysis. A more detailed discussion is given in the Supplementary Material.

\begin{table}[!th]
\centering
\scriptsize
\renewcommand{\arraystretch}{1}
\begin{tabular}{l|l|l}
\hline
Proporties              & Auto-Linear                     & Generative model-based \\ \hline
Paired Data             & Required                        & Not Required           \\ \hline
Inference               & Single Step                     & Single/Multiple Step             \\ \hline
Solution                & \begin{tabular}[c]{@{}l@{}}Condition Average \\  with Minimum Error\end{tabular} & Multiple Instances       \\ \hline
Forward Modeling        & Not Required                    & Required               \\ \hline
\begin{tabular}[c]{@{}l@{}}Solving \\ Forward Problem\end{tabular} & Can                             & Cannot                 \\ \hline
\end{tabular}
\vskip -0.1in
\caption{Differences Between Auto-Linear and Generative Model-Based Approaches.}
\label{table:Generative}
\vskip -0.25in
\end{table}

\section{Review of previous methods}
Let’s begin by reviewing the previous joint training method (InversionNet~\cite{wu2019inversionnet}) and the decouple training method (InvLInT~\cite{feng2022intriguing}). InversionNet considers FWI as an \textit{image-to-image translation problem}. As shown in Figure~\ref{teaser}-a, they train an encoder-decoder convolutional network to map seismic data $p(x, t)$ and velocity maps $c(x,z)$. This kind of jointly trained method can be formulated as:
\begin{align}
\theta^*, \eta^*&=\argmin_{\theta, \eta} \mathcal{L}(c, (\mathcal{D}_\eta \circ \mathcal{E}_\theta)(p)), \label{eq:invnet}
\end{align}
where $\mathcal{L}$ is the loss function. The encoder $\mathcal{E}$'s parameters $\theta$ and decoder $\mathcal{D}$'s parameters $\eta$ are jointly trained from the supervision of paired seismic data and velocity maps.

In InvLINT, the authors also try to decouple the encoder and decoder, and use a linear layer to connect two latent spaces. They use two pre-determined integral transforms, with Sine and Gaussian kernels, to embed the seismic data and velocity into high-dimensional spaces. They also provide a theoretical analysis with some hypotheses to establish a near-linear relationship in latent spaces when appropriate transforms are used. This method can be formulated as:
\begin{align}
&\bm{P} =[P_1, \dots, P_N]^T, \;\; P_n = \iint p(x, t)\Phi_n(x, t)dxdt, \nonumber\\
&\bm{C} =[C_1, \dots, C_M]^T, \;\; C_m = \iint c(x,z)\Psi_m(x,z)dxdz, \nonumber \\
&\bm{A}^*=\argmin_{\bm{A}}\mathcal{L}(\bm{C}, \bm{A}\bm{P}), \nonumber \\
&\eta^*=\argmin_{\eta} (c, (\mathcal{D}_\eta \circ \bm{A}^*)(\bm{P})),
\label{eq:linear-property}
\end{align}
where $\Phi_n$ and $\Psi_m$ are kernels for integral transforms.
Notably, InvLINT only decouples the training of encoders. Since there is no explicit inverse transformation of the Gaussian integral transform, the training of their decoder remains dependent on the seismic data $c$ for the specific dataset. 
Moreover, the kernel-based solution encounters significant limitations, including poor performance on datasets containing large variations and high-frequency components (e.g., OpenFWI~\cite{dengopenfwi}) and a lack of noise resistance and systematic rule for kernel selection tailored to different situations. Consequently, this pre-determined kernel solution might not suit wider applications.

\section{Methodology}
Expanding upon InvLINT's partially decoupled approach, our framework utilizes the advanced self-supervised learning strategy to decouple the training processes of the encoder and decoder completely, significantly enhancing performance beyond InvLINT's.
In this section, we present the formula of the Auto-Linear Phenomenon and describe how to apply it to the FWI problem.
Table~\ref{table:notations} lists used notations.

\begin{table}[]
\centering
\scriptsize
\renewcommand{\arraystretch}{1}
\begin{tabular}{l|l}
 \hline
Variable & Definition \\ \hline \hline
$p(x, t)$ &  seismic data      \\ \hline
$c(x, z)$         &  velocity maps          \\ \hline
$\mathcal{E}_s$ &  encoder of seismic data   \\ \hline
$\mathcal{D}_s$        &  decoder of seismic data   \\ \hline
$\mathcal{E}_v$        &  encoder of velocity map          \\ \hline
$\mathcal{D}_v$        &  decoder of velocity map          \\ \hline
$\bm{A}$, $\bm{B}$        &   Matrix         \\ \hline
\end{tabular}
\vskip -0.1in
\caption{Table of Notation.}
\label{table:notations}
\vskip -0.25in
\end{table}



\subsection{Auto-Linear Phenomenon}
Here, we define the Auto-Linear Phenomenon as the automatic integration of independently self-trained encoder and decoder from two different domains into an end-to-end model through a linear mapping. This phenomenon highlights two key principles: network integration and feature correlation. Network integration is facilitated by a linear mapping that effectively connects the independently trained encoder and decoder into a cohesive end-to-end model of subsurface imaging. This ensures the encoder and decoder operate independently yet synergistically within the broader framework. Moreover, the Auto-Linear Phenomenon reveals that the latent representations learned independently from each domain are inherently linearly correlated. 

Based on the above definition, the relationship of different components in Auto-Linear can be formulated as
\begin{align}
p&=(\mathcal{D}_s \circ \mathcal{E}_s)(p), \tag{4.1}\label{eq:2.1}\\ 
c&=(\mathcal{D}_v \circ \mathcal{E}_v)(c), \tag{4.2}\label{eq:2.2}\\
\mathcal{E}_v(c) &= \bm{A}\mathcal{E}_s(p), \tag{4.3}\label{eq:2.3}\\
\bm{B} \mathcal{E}_v(c) &= \mathcal{E}_s(p), \tag{4.4}\label{eq:2.4}
\end{align}
where $\bm{A}$ and $\bm{B}$ are the linear mappings, and not necessary to be full rank. In the above formulation, Eq. \ref{eq:2.1} and \ref{eq:2.2} describe two domain-independent autoencoders. Eq. \ref{eq:2.3} and \ref{eq:2.4} illustrate the linear correlations in the latent representations. Then, by plugging Eq. \ref{eq:2.3} into Eq. \ref{eq:2.2} (Eq. \ref{eq:2.4} into Eq. \ref{eq:2.1}), we can construct an inverse (forward) model, demonstrating a seamless integration of these processes.

In contrast, InvLINT focuses on the linear correlation between two domain embeddings. This results in a model that lacks the comprehensive encoder-decoder structure necessary for subsurface imaging. Consequently, InvLINT necessitates an additional training phase for a domain-dependent decoder, distinguishing it from the Auto-Linear.
Therefore, while InvLINT introduces the concept of linear correlation, its partial approach can be described as \textit{semi}-Auto-Linear because it only achieves linearity after encoding.

\subsection{Auto-Linear Framework in FWI}
Leveraging the above formulations, we introduce a novel approach in subsurface imaging that fully decouples the training of the encoder and decoder, enabling a more modular approach to model development in FWI. The training process can be formulated as:
\begin{align}
\theta_{s}^*, \eta_{s}^*&=\argmin_{\theta_{s}, \eta_{s}} \mathcal{L}(p, (\mathcal{D}_{s\eta_{s}} \circ \mathcal{E}_{s\theta_{s}})(p)), \tag{5.1}\label{eq:3.1}\\ 
\theta_{v}^*, \eta_{v}^*&=\argmin_{\theta_{v}, \eta_{v}} \mathcal{L}(c, (\mathcal{D}_{v\eta_{v}} \circ \mathcal{E}_{v\theta_{v}})(c)), \tag{5.2}\label{eq:3.2}\\
\bm{A}^*&=\argmin_{\bm{A}}\mathcal{L}(c, (\mathcal{D}_{v{\eta_{v}^*}} \circ \bm{A} \circ \mathcal{E}_{s{\theta_{s}^*}})(p)), \tag{5.3}\label{eq:3.3} \\ 
\bm{B}^*&=\argmin_{\bm{B}}\mathcal{L}(p, (\mathcal{D}_{s{\eta_{s}^*}} \circ \bm{B} \circ \mathcal{E}_{v{\theta_{v}^*}})(c)), \tag{5.4}\label{eq:3.4}
\end{align}
where $\mathcal{L}$ is the loss function. 
The entire model is trained in two steps: a domain-independent self-supervised learning step to train two autoencoders, and a supervised learning step to train the linear converters.
In the self-supervised learning step, as outlined in Eq. \ref{eq:3.1} and \ref{eq:3.2}, paired data are not needed. The encoders and decoders for seismic data, $\mathcal{E}_s$ and $\mathcal{D}_s$, and for velocity maps, $\mathcal{E}_v$ and $\mathcal{D}_v$, are trained independently using data from their respective domains. In the self-supervised learning step of the inverse process, as per \ref{eq:3.3}, the trained parameters of $\mathcal{E}_s$ and $\mathcal{D}_v$ are frozen. Subsequently, a linear converter $\bm{A}$ is trained using paired data to establish a connection between them.  This approach also applies to forward modeling, as indicated in \ref{eq:3.4}.

Although the seismic data is generated from the velocity,  the automatic linear correlation observed between the two learned representations is not trivial. This is because the parameters of seismic autoencoder $\mathcal{E}_s, \mathcal{D}_s$ are independent of the velocity $p$ and the parameters of velocity autoencoder $\mathcal{E}_v, \mathcal{D}_v$ are not influenced by the seismic $c$. Two MAEs are trained completely separately, and there is no inherent mechanism to guarantee that their respective learned representations will be naturally correlated.

Notably, due to a lack of constraint, the self-supervised learner can easily learn shortcuts for reconstruction. We chose to employ Masked Autoencoder (MAE)~\cite{he2022masked} as it generates better latent representations and can learn essential information about two physical quantities through the use of masks as noise, making it easier to connect the latent spaces of two modalities and transform and reconstruct the other. Furthermore, in practice, we decompose the linear converter into two linear layers with a low-dimensional bottleneck to constrain its rank. This effectively reduces redundancy in the network.

\subsection{Benefits of Auto-Linear Framework}

The Auto-Linear Framework introduces a range of benefits, spanning from its framework structure to the model property, and its practical performance.
At the framework level, it can simultaneously acquire encoder and decoder components for both forward and inverse problems. This eliminates the previous methods' need to retrain entirely new networks for different tasks, as only the linear layers require specific training.
From the perspective of model properties, the self-supervised pre-training captures essential information from both domains, granting encoders and decoders strong generalization abilities. This allows them to be effectively shared across datasets with various subsurface structures. 
In terms of actual performance, our model significantly outperforms InvLINT, especially in noise robustness. Compared to InversionNet, our model achieves comparable results, with only the linear layer needing paired data. Being smaller and having a simpler supervised training component, it requires less data and is less susceptible to overfitting. Thus, our model exhibits superior performance in few-shot scenarios and has improved noise robustness.

\section{Experiments}
We evaluate our approach on OpenFWI~\cite{dengopenfwi}, the first and only large-scale collection of openly accessible multi-structural seismic FWI datasets with benchmarks. We compare our method with the state-of-the-art works, including InversionNet~\cite{wu2019inversionnet}, i.e., the method that jointly trains the encoder and decoder, and InvLINT~\cite{feng2022intriguing}, i.e., the method that separates the encoder and decoder. We also evaluate Auto-Linear's generalizability for other imaging and PDE tasks. In particular, we test it on the electromagnetic (EM) inversion task controlled by Maxwell’s equations. In the Supplementary Material, we compare the latent representation learned by our method and InvLINT, evaluating the generalization ability of the encoder and decoder on a newly constructed dataset, discuss different factors that affect performance, and explore applying Auto-Linear to the elastic FWI, which involves multiple-input to multiple-output maps. For interested readers, \citet{dengopenfwi} provides a detailed comparison of the physics-driven method.

\subsection{Implementation Details} 
\textbf{Datasets.}
While real data are extremely expensive and difficult to obtain, subsurface imaging research often relies on full-physics simulations, driven by the lack of publicly available real datasets.
We verify our method on OpenFWI~\cite{dengopenfwi}, the first open-source collection of large-scale, multi-structural benchmark datasets for data-driven seismic FWI. It contains $11$ 2D datasets with baseline, which can be divided into four groups: four datasets in the ``Vel Family”, four datasets in the ``Fault Family”, two datasets in the ``Style Family”, and one dataset in the ``Kimberlina Family”. Four datasets in the ``Vel Family” are FlateVel-A/B, and CurveVel-A/B; four datasets in the ``Fault Family” are FlateFault-A/B, and CurveFault-A/B; two datasets in``Style Family” are Style-A/B; and one dataset in ``Kimberlina Family” is Kimberlina-CO$_2$. The first three families cover two versions: easy (A) and hard (B), in terms of the complexity of subsurface structures. We will use the abbreviations (e.g. FVA for FlatVel-A and CO$_2$ for Kimberlina-CO$_2$). More details can be found in~\cite{dengopenfwi}.

\textbf{Training Details.}
The input seismic data are normalized to the range [-1, 1] with a log scale. We employ AdamW \cite{loshchilov2018decoupled} optimizer with momentum parameters $\beta_1 = 0.9$, $\beta_2 = 0.999$ and a weight decay of $0.05$ for both self-supervision and supervision steps.
In the self-supervision step, we use the same hyper-parameters and the training schedule with the original MAE paper~\cite{he2022masked}, except we change the batch size to 512 and remove the pixel normalization.  We use each family together to train the MAE. Thus, in total, we trained four different models. 
In the supervision step, the initial learning rate is set to be $1 \times 10^{-3}$, and decayed with a cosine annealing \cite{loshchilov2016sgdr}. The batch size is set to 256. To make a fair comparison with the previous work, we use $l_1$ loss to train the linear layer. The exact network architectures are shown in Supplementary Material.
We implement our models in Pytorch and train them on 1 NVIDIA Tesla V100 GPU.

\textbf{Evaluation Metrics.}
We apply three metrics to evaluate the generated geophysical properties: MAE, MSE, and Structural Similarity (SSIM). Following the existing literature \cite{wu2019inversionnet, feng2022intriguing, dengopenfwi}, MAE and MSE are employed to measure the pixel-wise error, and SSIM is to measure the perceptual similarity since velocity has highly structured information, and degradation or distortion can be easily perceived by a human. 
We calculate them on normalized velocity maps, i.e., MAE and MSE in the scale $\left[-1, 1\right]$, and SSIM in the scale $\left[0, 1\right]$.

\begin{figure}[!t]
\begin{center}
\centerline{\includegraphics[height=.65\textheight]{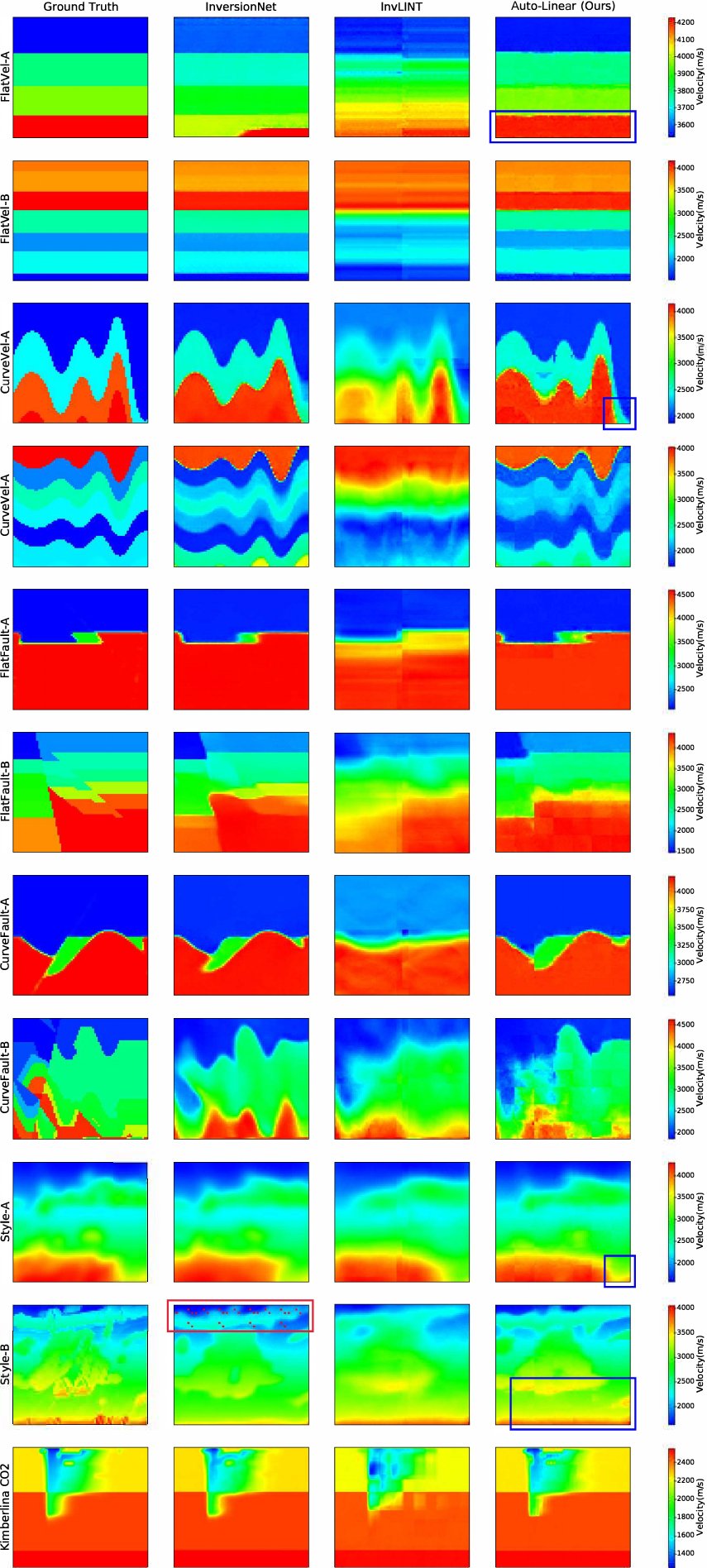}}
\vskip -0.1in
\caption{Illustration of results evaluated on OpenFWI, compared with InversionNet and InvLINT.} 
\label{main_result1}
\end{center}
\vskip -0.5in
\end{figure}

\begin{table*}[ht]
\renewcommand{\arraystretch}{1}
\centering
\scriptsize
\setlength{\tabcolsep}{2.9mm}
\begin{tabular}{c|l|c|c|c|c|c|c|c|c|c|c|c}
\thickhline
\multicolumn{1}{l|}{Metrics} &
  Model &
  FVA &
  FVB &
  CVA &
  CVB &
  FFA &
  FFB &
  CFA &
  CFB &
  SA &
  SB &
  CO2 \\ \thickhline
\multirow{2}{*}{MAE$\downarrow$} &
  Auto-Linear &
  \textbf{0.0081} &
  \underline{0.0467} &
  \underline{0.0738} &
  \underline{0.1820} &
  \textbf{0.0164} &
  \underline{0.1208} &
  \underline{0.0277} &
  \underline{0.1791} &
  \underline{0.0719} &
  \textbf{0.0638} &
  \textbf{0.0060} \\ \cline{2-13} 
 &
  InversionNet &
  \underline{0.0131} &
  \textbf{0.0351} &
  \textbf{0.0685} &
  \textbf{0.1497} &
  \underline{0.0172} &
  \textbf{0.1055} &
  \textbf{0.0260} &
  \textbf{0.1646} &
  \textbf{0.0625} &
  \underline{0.0689} &
  \underline{0.0061} \\ \cline{2-13} 
 &
  InvLINT &
  0.0532 &
  0.1621 &
  0.0981 &
  0.2462 &
  0.0729 &
  0.1522 &
  0.0853 &
  0.1955 &
  0.1002 &
  0.0835 &
  0.0150 \\ \thickhline
\multirow{2}{*}{MSE$\downarrow$} &
  Auto-Linear &
  \underline{0.0005} &
  \underline{0.0151} &
  \underline{0.0188} &
  \underline{0.1051} &
  \underline{0.0026} &
  \underline{0.0362} &
  \underline{0.0061} &
  \underline{0.0697} &
  \underline{0.0139} &
  \textbf{0.0097} &
  \underline{0.0017} \\ \cline{2-13} 
 &
  InversionNet &
  \textbf{0.0004} &
  \textbf{0.0077} &
  \textbf{0.0162} &
  \textbf{0.0836} &
  \textbf{0.0018} &
  \textbf{0.0303} &
  \textbf{0.0042} &
  \textbf{0.0614} &
  \textbf{0.0105} &
  \underline{0.0260} &
  \textbf{0.0014} \\ \cline{2-13} 
 &
  InvLINT &
  0.0085 &
  0.0650 &
  0.0238 &
  0.1312 &
  0.0190 &
  0.0467 &
  0.0229 &
  0.0754 &
  0.0209 &
  0.0132 &
  0.0039 \\ \thickhline
\multirow{2}{*}{SSIM$\uparrow$} &
  Auto-Linear &
  \underline{0.9888} &
  \underline{0.9044} &
  \underline{0.8057} &
  \underline{0.6169} &
  \underline{0.9701} &
  \underline{0.6868} &
  \underline{0.9426} &
  \underline{0.5672} &
  \underline{0.8423} &
  \textbf{0.7275} &
  \textbf{0.9908} \\ \cline{2-13} 
 &
  InversionNet &
  \textbf{0.9895} &
  \textbf{0.9461} &
  \textbf{0.8074} &
  \textbf{0.6727} &
  \textbf{0.9766} &
  \textbf{0.7208} &
  \textbf{0.9566} &
  \textbf{0.6136} &
  \textbf{0.8859} &
  \underline{0.6314} &
  \underline{0.9872} \\ \cline{2-13} 
 &
  InvLINT &
  0.8457 &
  0.6465 &
  0.7355 &
  0.4946 &
  0.8506 &
  0.6445 &
  0.8204 &
  0.5471 &
  0.7916 &
  0.6557 &
  0.9760 \\ \hline
\end{tabular}
\vskip -0.1in
\caption{Quantitative results evaluated on OpenFWI, compared with InversionNet and InvLINT, in terms of MAE, MSE, and SSIM.
Auto-Linear achieves comparable accuracy with InversionNet and outperforms InvLINT in terms of all three metrics. For each dataset, we use bold to highlight the best results, and underlined for the second best results.}
\label{table:main_result}
\vskip -0.2in
\end{table*}

\subsection{Auto-Linear is Simple and Effective in both Inverse and Forward Problem}
\textbf{Comparisons with the Joint Training Method.}
Table~\ref{table:main_result} shows the comparison results with InversionNet~\cite{wu2019inversionnet}. The results of InversionNet are the reported benchmark in~\cite{dengopenfwi}. Compared to the InversionNet, our Auto-Linear achieves comparable results on multiple datasets with only half the model size (12.3M vs. 24.4M), and only needs to supervised train the linear layer. In FlatVel-A/B, Style-B, and Kimberlina-CO$_2$, Auto-Linear even outperforms InversionNet in some metrics. The velocity maps inverted by different methods are shown in Figure~\ref{main_result1}. We can find InversionNet has a clearer boundary, while Auto-Linear is better at capturing the structure details in deep position (e.g., as boxed out on FaltVel-A, CurveVel-A, Style-B, and Style-A).
The corresponding error map and more visualizations are provided in the Supplementary Material. Note that, InversionNet in Style-B always outputs a strange pattern in results as boxed out in red.

\textbf{Comparisons with the Separate Training Method.}
We compare Auto-Linear with InvLINT~\cite{feng2022intriguing}, which also separates the encoder and decoder, and has a linear converter. Results are shown in Table~\ref{table:main_result}. Compared to InvLINT, Auto-Linear outperforms it in terms of all three metrics.
The velocity maps inverted by different methods are shown in Figure~\ref{main_result1}. The corresponding error map and more visualization results are provided in the Supplementary Material. 
We can clearly observe that InvLINT performs poorly for data with high-frequency layering locations and faults (i.e., ``Vel Family" and ``Fault Family"), but yields good results in smoother structures like ``Style Family" and Kimberlina CO$_2$.
This phenomenon may come from: 1) InvLINT model is very small and has limited expressive power. ``Vel Family" and ``Fault Family" are very diverse. It does not have enough capacity to learn all cases. 2) The Gaussian kernel cannot capture the small fault structure well, such as the interface and fault structures. 3) Their encoder uses frequency domain features. However, the high-frequency signal is mainly present in the reflected wave, which has a small amplitude. It is not easy to be captured by a frequency-domain encoder. A comparison of the seismic and velocity latent representations obtained by our method and InvLINT is presented in the Supplementary Material.

\begin{figure}[!ht]
\begin{center}
\centerline{\includegraphics[width=0.7\columnwidth]{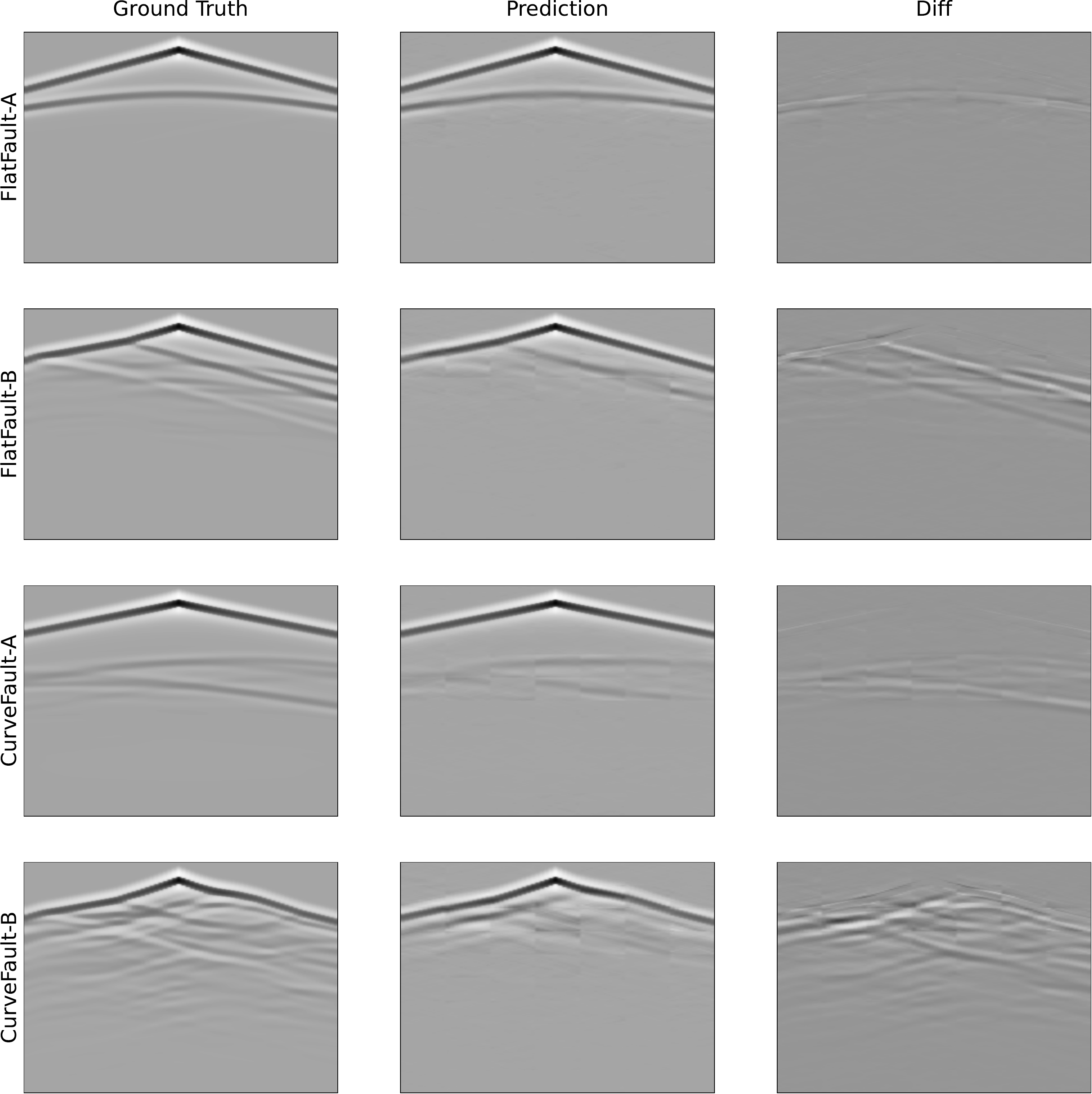}}
\vskip -0.1in
\caption{Illustration of forward results on ``Fault Family".} 
\label{forward_result}
\end{center}
\vskip -0.35in
\end{figure}
\begin{figure}[!ht]
\begin{center}
\centerline{\includegraphics[width=0.7\columnwidth]{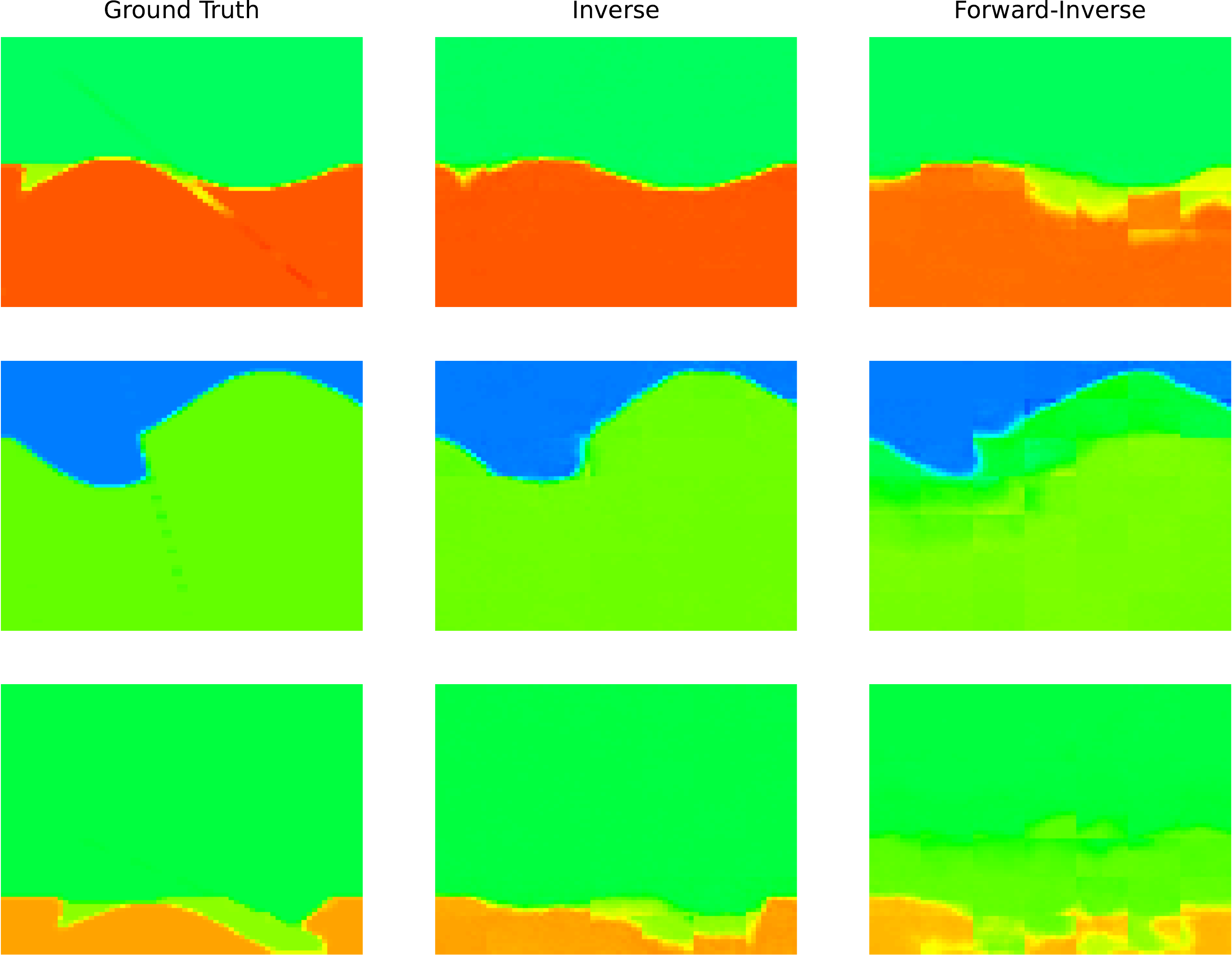}}
\vskip -0.1in
\caption{Illustration of results from Forward-Inverse process on CurveFault-A.} 
\label{forward_inverse}
\end{center}
\vskip -0.4in
\end{figure}

\textbf{Auto-Linear for Forward Process.} We further evaluate the efficacy of Auto-Linear for the forward process on ``Fault Family". Utilizing the pre-trained velocity encoder and seismic decoder, we trained a linear converter to map the velocity latent vector to the seismic latent vector. We combine $l_1$ and $l_2$ loss as the loss function, and compare our method with a forward version InversionNet \cite{gupta2023solving}. The results, detailed in Table~\ref{table:forward}, show a promising improvement in seismic data construction compared to InversionNet. We illustrate our output in Figure~\ref{forward_result}. 
While the reconstruction of reflected waves, particularly in more complex datasets, shows potential for further improvement, the overall capability to construct seismic data is evident.
Additionally, we conducted experiments using the trained inversion network above to inverse the generated seismic data by the forward model on CurveFault-A, and the visualizations are shown in Figure~\ref{forward_inverse}. Although the performance of this ``forward-inverse" model is diminished due to the cascade amplification of errors, the visualizations affirm that the recovered velocity maps successfully capture the main structures.

\begin{table}[h]
\renewcommand{\arraystretch}{1}
\centering
\scriptsize
\setlength{\tabcolsep}{1.8mm}
\begin{tabular}{c|c|c|c|c}
\thickhline
Dataset  & Model  & MAE$\downarrow$    & MSE$\downarrow$    & SSIM$\uparrow$   \\ \thickhline
\multirow{2}{*}{\begin{tabular}[c]{@{}l@{}}FlatFault-A\end{tabular}} & Auto-Linear & \textbf{0.0099} & \textbf{0.0006} & \textbf{0.9853} \\ \cline{2-5} 
                                                                     & InversionNet & 0.0332 & 0.0145 & 0.9659 \\ \hline
\multirow{2}{*}{\begin{tabular}[c]{@{}l@{}}FlatFault-B\end{tabular}} & Auto-Linear & \textbf{0.0193} & \textbf{0.0016} & \textbf{0.9604} \\ \cline{2-5} 
                                                                     & InversionNet & 0.0397 & 0.0078 & 0.9283 \\ \hline
\multirow{2}{*}{\begin{tabular}[c]{@{}l@{}}CurveFault-A\end{tabular}} & Auto-Linear & \textbf{0.0132} & \textbf{0.0009} & \textbf{0.9784} \\ \cline{2-5} 
                                                                     & InversionNet & 0.0373 & 0.0186 & 0.9571 \\ \hline 
\multirow{2}{*}{\begin{tabular}[c]{@{}l@{}}CurveFault-B\end{tabular}} & Auto-Linear & \textbf{0.0253} & \textbf{0.0022} & \textbf{0.9404} \\ \cline{2-5} 
                                                                     & InversionNet & 0.0594 & 0.0132 & 0.8636 \\ \hline
\end{tabular}
\vskip -0.1in
\caption{Quantitative results for the forward process.}
\label{table:forward}
\vskip -0.2in
\end{table}

\subsection{Auto-Linear is applicable to another subsurface imaging task.}
We experimented on another subsurface imaging task, recovering subsurface conductivity from surface-acquired electromagnetic (EM) measurements, on the Kimberlina-Reservoir dataset \cite{Development-2021-Alumbaugh, feng2022intriguing}. Let $\mathbf{E}$ and $\mathbf{H}$ are the electric and magnetic fields. $\mathbf{J}$ and $\mathbf{M}$ are the electric and magnetic sources. $\sigma$ is the electrical conductivity and $\mu_{0}=4\pi\times10^{-7} \Omega\cdot s/m$ is the magnetic permeability of free space. The governing equations here are Maxwell’s Equations 
\begin{align}
\sigma\mathbf{E}-\nabla\times\mathbf{H}&=-\mathbf{J},  \notag \\
\nabla\times\mathbf{E}+i\omega\mu_{0}\mathbf{H}&=-\mathbf{M}. \tag{5}
\label{eq:EMForward}
\end{align}
We compared the results of our Auto-Linear model with InversionNet and results reported in InvLINT (Feng et al., 2022), presented in Table~\ref{table:em}. Note that, to maintain consistency with InvLINT, the MAE and MSE reported below were calculated after denormalizing to the original range of $[0,0.65]$. For all other results presented in our paper, the MAE and MSE were calculated in the normalized range of $[-1,1]$. We observe that our proposed Auto-Linear yields significantly better performance than those obtained using InvLINT and InversionNet.

\begin{table}[th]
\renewcommand{\arraystretch}{1}
\centering
\scriptsize
\setlength{\tabcolsep}{1.8mm}
\begin{tabular}{c|c|c|c|c}
\thickhline
Dataset &      Model        & MAE$\downarrow$     & MSE$\downarrow$      & SSIM$\uparrow$   \\ \thickhline
\multirow{3}{*}{\begin{tabular}[c]{@{}l@{}}Kimberlina-\\ Reservoir\end{tabular}} & Auto-Linear & \textbf{0.00438} & \textbf{0.000192} & \textbf{0.9700} \\ \cline{2-5} 
        & InversionNet & 0.01330 & 0.000855 & 0.9175 \\ \cline{2-5} 
        & InvLINT      & \underline{0.00703} & \underline{0.000537} & \underline{0.9370} \\ \hline
\end{tabular}
\vskip -0.1in
\caption{Quantitative results for EM inversion. MAE and MSE are calculated after denormalizing to their original range ($[0,0.65]$). Highlighting the best results with bold and the second best results with underline.}
\label{table:em}
\vskip -0.2in
\end{table}

\subsection{Auto-Linear has Nice Properties}
In this part, we demonstrate that our Auto-Linear has some nice properties, including the strong generalization ability of the pre-trained encoder/decoder, good noise handling, solid performance on few-shot learning, and a correlation between linear layers and the datasets' complexity.

\textbf{Generalization Ability of Encoder and Decoder.}
We study the generalization ability of the pre-trained encoder and decoder. In particular, we choose the seismic encoder and velocity decoder that self-supervised trained on ``Fault Family", fix it, and train the linear converter on other datasets (except the Kimberlina-CO$_2$, since it has different dimensions). The results are shown in Table~\ref{table:share_en_de}.


\begin{table}[]
\renewcommand{\arraystretch}{1}
\centering
\scriptsize
\setlength{\tabcolsep}{1.4mm}
\begin{tabular}{c |l |c |c |c |c |c |c}
\thickhline
Metrics               & Model &       FVA & FVB & CVA & CVB & SA & SB  \\ \thickhline
                      & Auto-Linear       & \textbf{0.0073} & 0.0570    & \textbf{0.0653} & \textbf{0.1804} & 0.0725  & 0.0646  \\ \cline{2-8} 
\multirow{-2}{*}{MAE$\downarrow$} & InversionNet & 0.0131          & 0.0351    & 0.0685          & 0.1497          & 0.0625  & 0.0689  \\ \thickhline
                      & Auto-Linear       & 0.0005          & 0.0198    & \textbf{0.0159} & \textbf{0.1030}  & 0.0144  & 0.0099  \\ \cline{2-8} 
\multirow{-2}{*}{MSE$\downarrow$} & InversionNet & 0.0004          & 0.0077    & 0.0162          & 0.0836          & 0.0105  & 0.0260  \\ \thickhline
                      & Auto-Linear       & \textbf{0.9895} & 0.8752    & \textbf{0.8192} & 0.6044          & 0.8351  & 0.7222  \\ \cline{2-8} 
\multirow{-2}{*}{SSIM$\uparrow$} & InversionNet & 0.9895 & 0.9461 & 0.8074 & 0.6727 & 0.8423 & 0.7275 \\ \hline
\end{tabular}
\vskip -0.1in
\caption{Generalizability of pre-trained encoder and decoder, using InversionNet as a baseline.}
\label{table:share_en_de}
\vskip -0.2in
\end{table}

Results show that encoders and decoders trained on the "Fault Family" excel across datasets, including Style-A and Style-B, despite their distinct subsurface structures. Notably, when applied to simpler datasets like the "Vel Family," they outperform the results trained exclusively on these datasets, as highlighted in Table~\ref{table:main_result}. This demonstrates that the latent representations from self-supervision capture essential information transferable across datasets, and strategic selection of self-supervision data can enhance our method's performance. The current choice of using each family together in the self-supervision is not to lose generality

\begin{table*}[!ht]
\centering
\scriptsize
\setlength{\tabcolsep}{0.7mm}
\renewcommand{\arraystretch}{1}
\renewcommand{\bfdefault}{sb}
\begin{tabular}{l|ccc|ccc|ccc|ccc|ccc}
 \thickhline
\multirow{3}{*}{Model} &
  \multicolumn{3}{c|}{$\sigma^2$ =0} &
  \multicolumn{3}{c|}{$\sigma^2$ =1e-5} &
  \multicolumn{3}{c|}{$\sigma^2$ =5e-5} &
  \multicolumn{3}{c|}{$\sigma^2$ =1e-4} &
  \multicolumn{3}{c}{$\sigma^2$ =5e-4} \\
 &
  \multicolumn{3}{l|}{} &
  \multicolumn{3}{c|}{PSNR=70.49dB} &
  \multicolumn{3}{c|}{PSNR=63.48dB} &
  \multicolumn{3}{c|}{PSNR=60.45dB} &
  \multicolumn{3}{c}{PSNR=53.39dB} \\ \hline
 &
  MAE$\downarrow$ &
  MSE$\downarrow$ &
  SSIM$\uparrow$ &
  MAE$\downarrow$ &
  MSE$\downarrow$ &
  SSIM$\uparrow$ &
  MAE$\downarrow$ &
  MSE$\downarrow$ &
  SSIM$\uparrow$ &
  MAE$\downarrow$ &
  MSE$\downarrow$ &
  SSIM$\uparrow$ &
  MAE$\downarrow$ &
  MSE$\downarrow$ &
  SSIM$\uparrow$ \\ \thickhline
Auto-Linear &
  0.0277 &
  0.0061 &
  0.9426 &
  0.0354 &
  0.0070 &
  0.9387 &
  0.0508 &
  0.0102 &
  0.9255 &
  0.0630 &
  0.0139 &
  0.9113 &
  0.1093 &
  0.0339 &
  0.8308 \\
Degradation (\%) &
  \textbackslash{} &
  \textbackslash{} &
  \textbackslash{} &
  -27.80 &
  -14.75 &
  -0.41 &
  -83.39 &
  -67.21 &
  -1.81 &
  -127.44 &
  -127.87 &
  -3.32 &
  -294.58 &
  -455.74 &
  -11.86 \\ \hline
InversionNet &
  \multicolumn{1}{l}{0.0260} &
  \multicolumn{1}{l}{0.0042} &
  \multicolumn{1}{l|}{0.9566} &
  0.0332 &
  0.0050 &
  0.9539 &
  0.0696 &
  0.0133 &
  0.9290 &
  0.1439 &
  0.0479 &
  0.8830 &
  0.4496 &
  0.3948 &
  0.6407 \\
Degradation (\%) &
  \textbackslash{} &
  \textbackslash{} &
  \textbackslash{} &
  -27.69 &
  -19.05 &
  -0.28 &
  -167.69 &
  -216.67 &
  -2.89 &
  -453.46 &
  -4.57 &
  -7.69 &
  -1629.23 &
  -9300.00 &
  -33.02 \\ \hline
InvLINT &
  \multicolumn{1}{l}{0.0853} &
  \multicolumn{1}{l}{0.0229} &
  \multicolumn{1}{l|}{0.8204} &
  3.1849 &
  19.3293 &
  0.0449 &
  7.4442 &
  103.2302 &
  0.0172 &
  10.1643 &
  185.8730 &
  0.0084 &
  23.8050 &
  1033.9167 &
  0.0025 \\
Degradation (\%) &
  \textbackslash{} &
  \textbackslash{} &
  \textbackslash{} &
  -3633.76 &
  -84307.42 &
  -94.53 &
  -8627.08 &
  -450686.90 &
  -97.90 &
  -11815.94 &
  -22653.60 &
  -98.98 &
  -27807.39 &
  -4514820.09 &
  -99.70 \\ \hline
\end{tabular}
\vskip -0.1in
\caption{Quantitative results on CurveFault-A with Gaussian noise of varying variance $\sigma^2$ added during testing.}
\label{table:noise_testing}
\vskip -0.2in
\end{table*}

To further show the generalization ability of the pre-trained encoder and decoder and the performance improvement by picking self-supervision data, we conduct another experiment that trains MAE on cross-family datasets. In particular, CurveVel-A, FlatFault-A, and CurveFault-A are used. We test this pair of encoder and decoder in all datasets. Moreover, we constructed a new dataset from Marmousi, which is a completely different dataset from OpenFWI, to support the claim of strong generalization. Both results are in Supplementary Material.

\textbf{Good handling of noise.}
We provide the quantitative results of the robustness test. In particular, we add Gaussian Noise with different variances to the input seismic data during testing. The noise level is chosen in accordance with the previous work~\cite{jin2021unsupervised}. Table~\ref{table:noise_testing} shows the performance on CurveFault-A. We also include the noise's variance ($\sigma^2)$ and average peak-to-noise ratio (PSNR) in the table. PSNR of a sample is defined as 
\begin{equation}
    \text{PSNR} = 10\log_{10}\frac{(p_{max} - p_{min})^2}{\ell_2(p - p')}, \tag{4}
\end{equation}
where $p_{max}$ and $p_{min}$ denote the maximum and minimum possible values of the seismic data in a dataset, $p$ is the clean seismic data, and $p'$ is the noisy data. 

Compared to other models, our Auto-Linear is the most robust one to noise. The robustness of Auto-Linear shows in two aspects. First, its performance degradation on noisy data is smaller than others. Second, when the noise's variance is large ($\sigma^2 \geq$ 5e-5), our method outperforms InversionNet. 
This enhanced robustness can be attributed to the smaller size of our model and its simpler supervised training component, which together increase its robustness.
As expected, InvLINT is extremely sensitive to the noise, as it only uses a Fourier transform as its encoder.

\textbf{Strong Performance on Few-Shot Learning.}
One of the most important benefits of our method is it does not need paired data to train its encoder and decoder. Thus, we test Auto-Linear on the few-shot learning situation, where only a limited number of paired data exists, and compare it InversionNet. We chose five datasets as examples and tested the situation that only 1/10 or 1/20 paired data can be used in supervised learning. MAE results are reported in Table~\ref{table:less_data}. 

Across all datasets, our method consistently surpasses InversionNet's performance as the amount of paired data decreases, regardless of whether our model performs better (e.g., FlatVel-A and FlatFault-A) or not as well (e.g., CurveVel-A, CurveFault-A, and Style-A) compared to InversionNet with the full dataset. 
Additionally, the EM inversion on the Kimberlina-Reservoir dataset, detailed in \ref{table:em}, is another example of few-shot learning. Unlike the "Fault Family" or "Style Family," which includes 48k training data, the EM dataset comprises only 750 training samples. In this scenario, our model significantly outperforms InversionNet. Collectively, these outcomes highlight a robust advantage of our Auto-Linear approach in few-shot learning scenarios.

\begin{table}[ht]
\centering
\scriptsize
\setlength{\tabcolsep}{2.4mm}
\renewcommand{\arraystretch}{1}
\renewcommand{\bfdefault}{sb}
\begin{tabular}{l|l|c|c|c}
\hline
Dataset & Model & \multicolumn{1}{l|}{Ratio=1} & \multicolumn{1}{l|}{Ratio=1/10} & \multicolumn{1}{l}{Ratio=1/20} \\ \hline
\multirow{2}{*}{FlatVel-A}    & Auto-Linear       & \textbf{0.0081} & \textbf{0.0361} & \textbf{0.0570} \\ \cline{2-5} 
                              & InversionNet & 0.0131          & 0.0590          & 0.0795          \\ \hline
\multirow{2}{*}{CurveVel-A}   & Auto-Linear       & 0.0738          & \textbf{0.1245} & \textbf{0.1444} \\ \cline{2-5} 
                              & InversionNet & \textbf{0.0685} & 0.1295          & 0.1450          \\ \hline
\multirow{2}{*}{FlatFault-A}  & Auto-Linear       & \textbf{0.0164} & \textbf{0.0423} & \textbf{0.0592} \\ \cline{2-5} 
                              & InversionNet & 0.0172          & 0.0544          & 0.0775          \\ \hline
\multirow{2}{*}{CurveFault-A} & Auto-Linear       & 0.0277          & \textbf{0.0634} & \textbf{0.0836} \\ \cline{2-5} 
                              & InversionNet & \textbf{0.0260} & 0.0781          & 0.1032          \\ \hline
\multirow{2}{*}{Style-A}      & Auto-Linear       & 0.0719          & \textbf{0.0989} & \textbf{0.1101} \\ \cline{2-5} 
                              & InversionNet & \textbf{0.0625} & 0.1046          & 0.1292     \\ \hline
\end{tabular}
\vskip -0.1in
\caption{MAE quantitative results using partial data sets. Ration indicates the proportion of data sets used.}
\label{table:less_data}
\vskip -0.1in
\end{table}

\begin{figure}[!t]
\begin{center}
\centerline{\includegraphics[width=0.6\columnwidth]{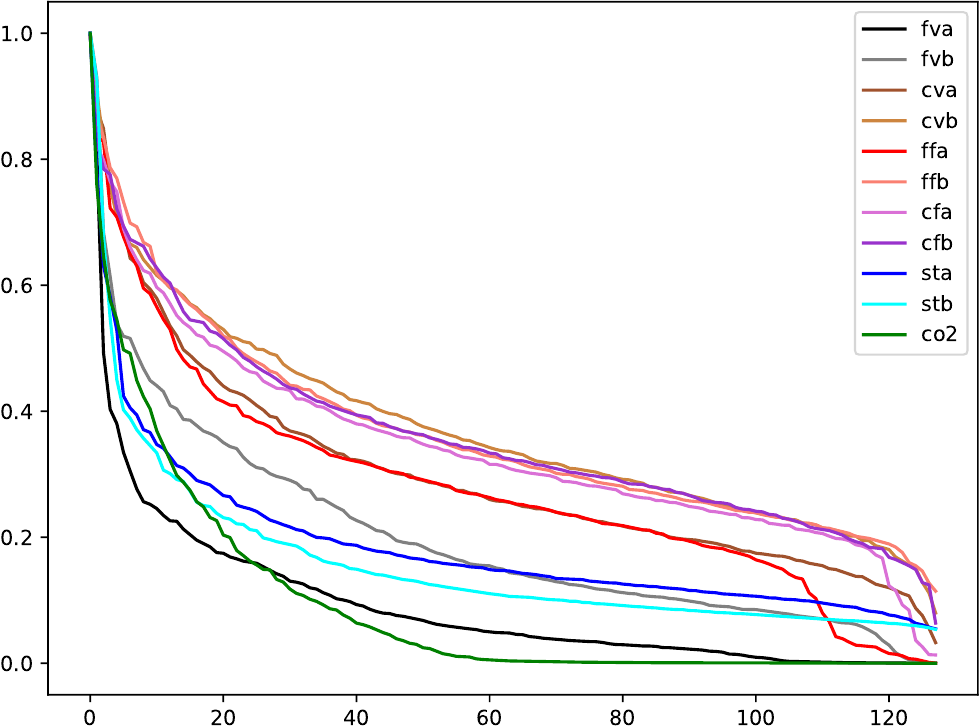}}
\caption{Normalized Singular Value of the linear layers.} 
\label{svd}
\end{center}
\vskip -0.45in
\end{figure}

\textbf{Correlation between the linear layer and datasets' complexity.}
By simplifying the image-to-image translation problem to a linear problem, our model is easy to analyze. With only a linear converter trained in a supervised manner, we can conduct a singular value decomposition analysis. The results are shown in Figure~\ref{svd}. We normalize it by dividing it by its maximum value and trunk it at 128 dim, which is the bottleneck dimension. We can clearly observe a strong correlation between the singular values and the dataset complexity. Generally speaking, CurveFault-B, FlatFault-B, Style-A, Style-B, and CurveFault-B have the most complex velocity map among all datasets. Their singular values are much slower to fall. On the other hand, FlatVel-A and Kimberlina-CO$_2$ are the simplest datasets, which are also reflected in their singular values. The original singular value is in the Supplementary Material for reference. Results prove that our linear converter correlates to the datasets' complexity.
For readers who might be interested, \citet{dengopenfwi} provide a more detailed analysis of datasets' complexity.

\subsection{Ablation Test}
In this part, we test the performance of several different non-linear converters, showing that they can only provide limited improvement. 
We also show the comparison among different self-supervised training methods, demonstrating our framework is adaptable beyond MAE to include other advanced self-supervised learners, as they impose sufficient constraints. We then demonstrate the influence of the local linear relationship between two latent spaces and test how the model's hyper-parameters (e.g., the rank of the linear converter) will influence the performance. The detailed results are shown in the Supplementary Material.

\section{Discussion}

\textbf{\quad Limitation about the collection of real data. }
Seismic surveys for collecting real data for FWI typically involve deploying an array of sensors, such as geophones or hydrophones, to record reflected and refracted seismic waves generated by controlled sources. 
However, the financial and computational demands, alongside the need for expert analysis, make acquiring extensive real datasets for supervised learning extremely difficult and expensive. To our knowledge, no open-source real-data dataset is available in seismic imaging. Due to the lack of labeled real datasets in subsurface geophysics, it is hard to train and evaluate our model on real data.

Simulated data plays a crucial role in this area's research, offering large-scale, clear datasets without noise. This is vital for analyzing complex relationships between seismic data and velocity maps. Despite challenges in training models solely on real data due to the lack of labeled real data in subsurface geophysics, this is a common limitation in the seismic inversion community.
The “Sim2Real” technique is well-received to transfer knowledge learned in simulation to real data \cite{james2019sim}. To mitigate the gap between simulation and real scenarios, we also have tested our model in velocity maps that yield physically realistic subsurface structures, i.e., Style-A and Style-B \cite{feng2021multiscale}. Additionally, we have imposed noise to simulate more realistic measurement procedures. Our method demonstrated promising performance in both scenarios. We will explore how to train the converter with purely unpaired data and mitigate the knowledge gap of real data in our future work.

\textbf{Fixed grid solutions.}
Our approach, along with many other data-driven methods using image-to-image translation or generative methods, duel the problem with a fixed grid. This contrasts with some physical-driven methods, which treat physical quantities as continuous functions. While physics-driven methods theoretically offer infinite resolution, they often produce overly smooth, low-resolution results in practice and struggle with large data sets and inter-dataset relationships. Conversely, although the computer vision method fixes the grid, at this fixed resolution, they can always achieve high-resolution results that surpass the physical method. The operator learning method presents a grid-free machine learning method; however, it still faces limitations (e.g., the requirements for the coverage of the input space, the smoothness of the output space, etc.), and the grid-based method is still the most adopted one of many scientific problems in both data-driven methods 
and traditional numerical solutions for PDEs.

\section{Conclusion}
We introduce a novel Auto-Linear Phenomenon in subsurface imaging, which describes the automatic emergence of a linear correlation between the latent spaces of two self-supervised autoencoders trained independently in different domains. Based on the phenomenon, we present a new framework that decouples the encoder's and decoder's training and simplifies the problem from image-to-image translation into a linear problem. In experiments, Auto-Linear achieved comparable performance and showed solid performance in a few-shot situation and robustness test.

\section*{Impact Statement}
This paper presents work whose goal is to advance the field of Machine Learning. There are many potential societal consequences of our work, none which we feel must be specifically highlighted here.

\bibliography{example_paper}

\begin{thebibliography}{23}
\providecommand{\natexlab}[1]{#1}
\providecommand{\url}[1]{\texttt{#1}}
\expandafter\ifx\csname urlstyle\endcsname\relax
  \providecommand{\doi}[1]{doi: #1}\else
  \providecommand{\doi}{doi: \begingroup \urlstyle{rm}\Url}\fi

\bibitem[Alumbaugh et~al.(2021{\natexlab{a}})Alumbaugh, Commer, Crandall, Gasperikova, Feng, Harbert, Li, Lin, Manthila~Samarasinghe, and Yang]{Development-2021-Alumbaugh}
Alumbaugh, D., Commer, M., Crandall, D., Gasperikova, E., Feng, S., Harbert, W., Li, Y., Lin, Y., Manthila~Samarasinghe, S., and Yang, X.
\newblock Development of a multi-scale synthetic data set for the testing of subsurface {CO$_2$} storage monitoring strategies.
\newblock In \emph{American Geophysical Union~(AGU)}, 2021{\natexlab{a}}.

\bibitem[Alumbaugh et~al.(2021{\natexlab{b}})Alumbaugh, Commer, Crandall, Gasperikova, Feng, Harbert, Li, Lin, Samarasinghe, and Yang]{alumbaugh2021development}
Alumbaugh, D., Commer, M., Crandall, D., Gasperikova, E., Feng, S., Harbert, W., Li, Y., Lin, Y., Samarasinghe, S., and Yang, X.
\newblock Development of a multi-scale synthetic data set for the testing of subsurface co2 storage monitoring strategies.
\newblock In \emph{AGU Fall Meeting Abstracts}, volume 2021, pp.\  S25A--0212, 2021{\natexlab{b}}.

\bibitem[Araya-Polo et~al.(2018)Araya-Polo, Jennings, Adler, and Dahlke]{araya2018deep}
Araya-Polo, M., Jennings, J., Adler, A., and Dahlke, T.
\newblock Deep-learning tomography.
\newblock \emph{The Leading Edge}, 37\penalty0 (1):\penalty0 58--66, 2018.

\bibitem[Chen et~al.(2021)Chen, Tachella, and Davies]{chen2021equivariant}
Chen, D., Tachella, J., and Davies, M.~E.
\newblock Equivariant imaging: Learning beyond the range space.
\newblock In \emph{Proceedings of the IEEE/CVF International Conference on Computer Vision}, pp.\  4379--4388, 2021.

\bibitem[Chen et~al.(2023)Chen, Dai, Chen, Liu, Yuan, Liu, and Lin]{chen2023image}
Chen, Y., Dai, X., Chen, D., Liu, M., Yuan, L., Liu, Z., and Lin, Y.
\newblock Image is first-order norm+ linear autoregressive.
\newblock \emph{arXiv preprint arXiv:2305.16319}, 2023.

\bibitem[Deng et~al.(2022)Deng, Feng, Wang, Zhang, Jin, Feng, Zeng, Chen, and Lin]{dengopenfwi}
Deng, C., Feng, S., Wang, H., Zhang, X., Jin, P., Feng, Y., Zeng, Q., Chen, Y., and Lin, Y.
\newblock Openfwi: Large-scale multi-structural benchmark datasets for full waveform inversion.
\newblock 2022.

\bibitem[Feng et~al.(2021)Feng, Fu, Feng, and Schuster]{feng2021multiscale}
Feng, S., Fu, L., Feng, Z., and Schuster, G.~T.
\newblock Multiscale phase inversion for vertical transverse isotropic media.
\newblock \emph{Geophysical Prospecting}, 69\penalty0 (8-9):\penalty0 1634--1649, 2021.

\bibitem[Feng et~al.(2023)Feng, Wang, Deng, Feng, Liu, Zhu, Jin, Chen, and Lin]{feng2023efwi}
Feng, S., Wang, H., Deng, C., Feng, Y., Liu, Y., Zhu, M., Jin, P., Chen, Y., and Lin, Y.
\newblock $\mathbf{{E}^{FWI}}$: Multi-parameter benchmark datasets for elastic full waveform inversion of geophysical properties.
\newblock \emph{CoRR}, 2023.

\bibitem[Feng et~al.(2022)Feng, Chen, Feng, Jin, Liu, and Lin]{feng2022intriguing}
Feng, Y., Chen, Y., Feng, S., Jin, P., Liu, Z., and Lin, Y.
\newblock An intriguing property of geophysics inversion.
\newblock \emph{The Thirty-ninth International Conference on Machine Learning}, 2022.

\bibitem[Gupta(2023)]{gupta2023solving}
Gupta, N.
\newblock \emph{Solving Forward and Inverse Problems for Seismic Imaging using Invertible Neural Networks}.
\newblock PhD thesis, Virginia Tech, 2023.

\bibitem[He et~al.(2022)He, Chen, Xie, Li, Doll{\'a}r, and Girshick]{he2022masked}
He, K., Chen, X., Xie, S., Li, Y., Doll{\'a}r, P., and Girshick, R.
\newblock Masked autoencoders are scalable vision learners.
\newblock In \emph{Proceedings of the IEEE/CVF Conference on Computer Vision and Pattern Recognition}, pp.\  16000--16009, 2022.

\bibitem[James et~al.(2019)James, Wohlhart, Kalakrishnan, Kalashnikov, Irpan, Ibarz, Levine, Hadsell, and Bousmalis]{james2019sim}
James, S., Wohlhart, P., Kalakrishnan, M., Kalashnikov, D., Irpan, A., Ibarz, J., Levine, S., Hadsell, R., and Bousmalis, K.
\newblock Sim-to-real via sim-to-sim: Data-efficient robotic grasping via randomized-to-canonical adaptation networks.
\newblock In \emph{Proceedings of the IEEE/CVF Conference on Computer Vision and Pattern Recognition}, pp.\  12627--12637, 2019.

\bibitem[Jin et~al.(2022)Jin, Zhang, Chen, Huang, Liu, and Lin]{jin2021unsupervised}
Jin, P., Zhang, X., Chen, Y., Huang, S.~X., Liu, Z., and Lin, Y.
\newblock Unsupervised learning of full-waveform inversion: Connecting {CNN} and partial differential equation in a loop.
\newblock In \emph{Proceedings of the Tenth International Conference on Learning Representations~(ICLR)}, 2022.

\bibitem[Lin et~al.(2023)Lin, Theiler, and Wohlberg]{Physics-2023-Lin}
Lin, Y., Theiler, J., and Wohlberg, B.
\newblock Physics-guided data-driven seismic inversion: Recent progress and future opportunities in full waveform inversion.
\newblock \emph{IEEE Signal Processing Magazine}, 40:\penalty0 115--133, 2023.

\bibitem[Loshchilov \& Hutter(2016)Loshchilov and Hutter]{loshchilov2016sgdr}
Loshchilov, I. and Hutter, F.
\newblock Sgdr: Stochastic gradient descent with warm restarts.
\newblock \emph{arXiv preprint arXiv:1608.03983}, 2016.

\bibitem[Loshchilov \& Hutter(2018)Loshchilov and Hutter]{loshchilov2018decoupled}
Loshchilov, I. and Hutter, F.
\newblock Decoupled weight decay regularization.
\newblock In \emph{Sixth International Conference on Learning Representations~(ICLR)}, 2018.

\bibitem[Schuster(2017)]{schuster2017seismic}
Schuster, G.~T.
\newblock \emph{Seismic inversion}.
\newblock Society of Exploration Geophysicists, 2017.

\bibitem[Song et~al.(2021)Song, Shen, Xing, and Ermon]{song2021solving}
Song, Y., Shen, L., Xing, L., and Ermon, S.
\newblock Solving inverse problems in medical imaging with score-based generative models.
\newblock \emph{arXiv preprint arXiv:2111.08005}, 2021.

\bibitem[Sun et~al.(2021)Sun, Innanen, and Huang]{sun2021physics}
Sun, J., Innanen, K.~A., and Huang, C.
\newblock Physics-guided deep learning for seismic inversion with hybrid training and uncertainty analysis.
\newblock \emph{Geophysics}, 86\penalty0 (3):\penalty0 R303--R317, 2021.

\bibitem[Wang et~al.(2023)Wang, Huang, and Alkhalifah]{wang2023prior}
Wang, F., Huang, X., and Alkhalifah, T.~A.
\newblock A prior regularized full waveform inversion using generative diffusion models.
\newblock \emph{IEEE Transactions on Geoscience and Remote Sensing}, 61:\penalty0 1--11, 2023.

\bibitem[Wu \& Lin(2019)Wu and Lin]{wu2019inversionnet}
Wu, Y. and Lin, Y.
\newblock {InversionNet}: An efficient and accurate data-driven full waveform inversion.
\newblock \emph{IEEE Transactions on Computational Imaging}, 6:\penalty0 419--433, 2019.

\bibitem[Zeng et~al.(2021)Zeng, Feng, Wohlberg, and Lin]{zeng2021inversionnet3d}
Zeng, Q., Feng, S., Wohlberg, B., and Lin, Y.
\newblock Inversionnet3d: Efficient and scalable learning for 3-d full-waveform inversion.
\newblock \emph{IEEE Transactions on Geoscience and Remote Sensing}, 60:\penalty0 1--16, 2021.

\bibitem[Zhang et~al.(2019)Zhang, Wu, Zhou, and Lin]{zhang2019velocitygan}
Zhang, Z., Wu, Y., Zhou, Z., and Lin, Y.
\newblock Velocitygan: Subsurface velocity image estimation using conditional adversarial networks.
\newblock In \emph{2019 IEEE Winter Conference on Applications of Computer Vision (WACV)}, pp.\  705--714. IEEE, 2019.

\end{thebibliography}
\bibliographystyle{icml2024}

\newpage
\appendix
\onecolumn

\section{Appendix}

\subsection{Comparing with generative model-based methods}
An alternative self-supervised approach for inverse problems involves pre-training a generative model on physical properties (e.g., velocity model and medical image) to capture their prior distribution. Subsequently, the model is adapted by integrating measurements and a physical model of the measurement process into the sampling process~\cite{wang2023prior, song2021solving}. Table~\ref{table:Generative} presents a side-by-side comparison, highlighting differences between our approach and generative model-based approaches. We compared the Auto-liner and generative model-based methods in detail.

Firstly generative-model-based approaches can be trained in a purely self-supervised manner and eliminate the need for paired data. In contrast, our approach needs the paired data to train the linear converter. Generally, real data only contains one or a limited number of unlabeled data, which makes it challenging to train our approach but the generative model-based approaches can still work on that.
For example, in \citet{wang2023prior}, a diffusion model-based method, trained with OpenFWI, was successfully applied to real marine data. 

Secondly, our encoder-decoder architecture enables single-step inference, unlike generative model-based methods that usually require multiple-step denoising when the diffusion model is used. 

Thirdly, in addressing the solution of FWI problems, our method aims for a maximum likelihood solution, also referred to as conditional average with minimum error, based on the distribution of training data, which can only generate one of the potential solutions. This approach is distinct from generative models, which seek distribution-to-distribution mappings and inherently can generate multiple solutions. 

Additionally, generative models-based methods incorporate the forward modeling process into learning, which requires a thorough understanding of the governing physics. This approach can greatly benefit from the existing knowledge of the forward operator, particularly when forward modeling is accessible, as in the cases of  FWI, CT, and MRI. However, challenges arise in scenarios where the forward modeling is partially unknown, completely unknown, or lacks an explicit formulation. For instance, in the Kimberlina carbon sequestration problem~\cite{alumbaugh2021development}, while data, generated by Maxwell’s Equations, has been made available, the forward modeling remains proprietary and challenging to replicate. In such scenarios, our approach presents an effective alternative, offering a different way for inverse analysis. 

Moreover, for problems like FWI, forward modeling is usually computationally expensive and time-consuming, leading to inefficient inference. For real-world data challenges, generative model-based methods can create paired data, which can be further utilized to train our Auto-Linear model, thus enhancing inference speed. This collaboration allows the Auto-Linear model and generative approaches to complement each other effectively in practical applications.

Finally, it is also important to note that generative models-based methods, while relying on forward modeling, are currently focused on inverse problems. In contrast, Auto-Linear has the capability to solve both forward modeling and inverse problem, simultaneously.

In summary,  Auto-Linear and generative model-based methods each contribute uniquely to our understanding of FWI, offering insights from different perspectives. Both approaches shed light on distinct aspects of the problem and suggest diverse strategies for tackling the inherent challenges of the inverse problem.

\subsection{Architecture.}
The exact transformer architectures and layer dimensions of the seismic and velocity autoencoders are provided in Table~\ref{table:dimensions}. For all the datasets except Kimberlina, the size of seismic data is $1000\times70$ and the size of velocity maps is $70\times70$. We choose the patch size $100\times10$ for seismic data and $10\times10$ for velocity maps. Thus, the latent dimension of seismic data is $132\times70$, and the latent dimension of velocity maps is $516\times49$. For Kimberlina, the patch size of seismic data is $250\times10$, and of velocity maps is $20\times40$. The latent dimension of seismic data is $132\times50$, and the latent dimension of velocity maps is $516\times70$. The rank of the linear converter is set to $128$. The mask ratio for training MAE is set to $0.75$.

\begin{table}[!h]
\renewcommand{\arraystretch}{1.3}
\centering
\scriptsize
\setlength{\tabcolsep}{0.6mm}
\begin{tabular}{c|c|c|c|c}
\hline
Model            & \#Layers & Embedded Dim & MLP Dim & \#Heads \\ \hline
Seismic Encoder  & 2        & 132          & 528     & 12      \\ \hline
Seismic Decoder  & 2        & 512          & 144     & 16      \\ \hline
Velocity Encoder & 3        & 516          & 2064    & 12      \\ \hline
Velocity Decoder & 2        & 512          & 2064    & 16      \\ \hline
\end{tabular}
\caption{Details of seismic and velocity autoencoders}
\label{table:dimensions}
\vskip -0.2in
\end{table}

\subsection{Generalizability.}
In the following experiment, CurveVel-A, FlatFault-A, and CurveFault-A are used to pre-train the MAEs. We test the generalizability of this pair of encoder and decoder in all datasets. Results are shown in Table~\ref{table:share_en_de2}.
\begin{table}[ht]
\centering
\begin{tabular}{l|c|c|c}
\hline
Dataset       & MAE$\downarrow$    & MSE$\downarrow$    & SSIM$\uparrow$   \\ \hline
CurveVel-A*   & 0.0634 & 0.0155 & 0.8267 \\ \hline
FlatFault-A*  & 0.0166 & 0.0026 & 0.9698 \\ \hline
CurveFault-A* & 0.0271 & 0.006  & 0.9434 \\ \thickhline
FlatVel-A     & 0.0072 & 0.0004 & 0.9912 \\ \hline
FlatVel-B     & 0.0552 & 0.0179 & 0.8783 \\ \hline
CurveVel-B    & 0.1754 & 0.0981 & 0.6157 \\ \hline
FlatFault-B   & 0.1260  & 0.0381 & 0.6734 \\ \hline
CurveFault-B  & 0.1837 & 0.0711 & 0.5590  \\ \hline
Style-A       & 0.0744 & 0.0146 & 0.8311 \\ \hline
Style-B       & 0.0653 & 0.0102 & 0.7175 \\ \hline
\end{tabular}
\caption{Quantitative results of the generalization ability of pre-trained encoder and decoder. The encoder and decoder are trained across datasets' families. (*) indicates the datasets used to train the encoder and decoder.}
\label{table:share_en_de2}
\end{table}

To support our claim of strong generalization, we conducted another experiment by evaluating our encoder and decoder on a completely new dataset. Since there is no suitable dataset for training the model, we created a new dataset derived from the original Marmousi velocity map. The Marmousi velocity map is a standard and complex benchmark widely recognized in the exploration geophysics community. The size of the original Marmousi is [13601, 2801]. After removing the water layer at the top, we obtained a velocity map of size [13601, 2381]. Utilizing a $70\times70$ sliding window, we generated a dataset with 25,000 samples, referred to as the ``Marmousi slice''. Additionally, we constructed another downsampled version by first downsampling the raw velocity to [2197, 731] and then applying the sliding window to produce 1,000 samples, referred to as ``Marmousi downsamples'' in the following. This downsampled version contains fewer samples and encapsulates more complex structures within each sample. For both datasets, we allocated 80\% of the data for training the linear converter, with the remaining 20\% used for validation. We tested the encoder and decoder, pre-trained on the ``Fault Family'' dataset, and compared the performance against supervised InversionNet trained on these two datasets as baselines. The results in Table~\ref{table:marmousi} show that our pre-trained models achieve comparable performance to the supervised baselines, indicating a strong generalization ability.

\begin{table}[th]
\renewcommand{\arraystretch}{1}
\centering
\scriptsize
\setlength{\tabcolsep}{1.8mm}
\begin{tabular}{c|c|c|c|c}
\thickhline
Dataset &      Model        & MAE$\downarrow$     & MSE$\downarrow$      & SSIM$\uparrow$   \\ \thickhline
\multirow{3}{*}{Marmousi slice} & Auto-Linear & 0.0116 & 0.0013 & 0.9770 \\ \cline{2-5} 
        & InversionNet & \textbf{0.0101} & \textbf{0.0012} & \textbf{0.9832} \\ \hline
\multirow{3}{*}{Marmousi downsample} & Auto-Linear & \textbf{0.0971} & \textbf{0.0296} & 0.8020 \\ \cline{2-5} 
        & InversionNet & 0.1094 & \textbf{0.0296} & \textbf{0.8311} \\ \hline
\end{tabular}
\vskip -0.1in
\caption{Quantitative results for generalization ability on completely different datasets.}
\label{table:marmousi}
\vskip -0.2in
\end{table}

\subsection{Ablation Test.}
In this part, we test the performance of several different non-linear converters and demonstrate the influence of the local linear relationship between two latent spaces. We also show the comparison between using masked autoencoder and autoencoder as self-supervised learners; and test the performance of several different non-linear converters and how the rank of the linear converter will influence the performance.

\textbf{Non-Linear Converter.}
We evaluate networks with a more complicated nonlinear converter on CurveFault-A. We tested four different settings: 1) a two-layer MLP; 2) a two-piece Maxout layer; 3) a two-layer U-Net; and 4) a four-layer U-Net. The results are provided in Table~\ref{table:nolinear}. From the results, we can see that 1) a simple nonlinear mapping (e.g., two-layer MLP or U-Net) has no positive effect on final performance; and 2) a piece-wise linear mapping (Maxout) or a much more complex nonlinear mapping (four-layer U-Net) can only provide limited improvement. These results are consistent with our conclusion of a near-linear relationship.

\begin{table}[h]
\renewcommand{\arraystretch}{1}
\centering
\scriptsize
\setlength{\tabcolsep}{1.8mm}
\begin{tabular}{c|c|c|c}
\thickhline
Model                  & MAE$\downarrow$    & MSE$\downarrow$    & SSIM$\uparrow$   \\ \thickhline
Linear               & 0.0277 & 0.0061 & 0.9426 \\ \hline
Two-Layer MLP         & 0.0280 & 0.0064 & 0.9433 \\ \hline
Two-Pieces Maxout & 0.0260 & 0.0057 & 0.9472 \\ \hline
2-Layer U-Net           & 0.0285 & 0.0062 & 0.9414 \\ \hline
4-Layer U-Net           & 0.0259 & 0.0056 & 0.9465 \\ \hline
\end{tabular}
\vskip -0.1in
\caption{Quantitative results on CurveFault-A with different nonlinear converters.}
\label{table:nolinear}
\vskip -0.2in
\end{table}

\textbf{Local Linear Relationship.}
We demonstrate the property we found that each dataset shows linear relation locally, and there is a piece-wise linear relation globally over multiple datasets. In particular, we let datasets in each family share not only the encoder and decoder but also the linear converter. In other words, we use all datasets in each family to train the linear converter. We report the results and performance change in Table~\ref{table:share_linear}.
In the table, we highlight the improvement of the results after sharing the linear converter. It is quite interesting that, generally, the datasets with a more complex subsurface structure show a performance improvement. In contrast, simpler datasets' performance drops a lot. The results come from the fact that a complex dataset covers a larger range in the latent space. The scope of simple datasets is covered by those complex ones in the same family. Thus, with more data to use, Auto-Linear achieves better results on complex datasets. But, for simple datasets, out-of-distribution data make the learning results deviate substantially from their local linear relationship.

\begin{table*}[!t]
\renewcommand{\arraystretch}{1}
\centering
\scriptsize
\setlength{\tabcolsep}{1.4mm}
\begin{tabular}{c|l?c|c|c|c?c|c|c|c?c|c}
\thickhline
\multicolumn{1}{l|}{Metrics} & Model & FlatVel-A & FlatVel-B & CurveVel-A & CurveVel-B & FlatFault-A & FlatFault-B & CurveFault-A & CurveFault-B & Style-A & Style-B \\ \thickhline
\multirow{2}{*}{MAE$\downarrow$} &
  Original &
  0.0081 &
  0.0467 &
  0.0738 &
  0.1820 &
  0.0164 &
  0.1208 &
  0.0277 &
  0.1791 &
  0.0719 &
  0.0638 \\ \cline{2-12} 
 &
  Sharing Linear &
  0.0191 &
  0.0545 &
  0.0761 &
  \textbf{0.1709} &
  0.0306 &
  \textbf{0.1198} &
  0.0421 &
  \textbf{0.1697} &
  \textbf{0.0699} &
  \textbf{0.0636} \\ \thickhline
\multirow{2}{*}{MSE$\downarrow$} &
  Original &
  0.0005 &
  0.0151 &
  0.0188 &
  0.1051 &
  0.0026 &
  0.0362 &
  0.0061 &
  0.0697 &
  0.0139 &
  0.0097 \\ \cline{2-12} 
 &
  Sharing Linear &
  0.0015 &
  0.0161 &
  0.0193 &
  \textbf{0.0963} &
  0.0054 &
  \textbf{0.0339} &
  0.0093 &
  \textbf{0.063} &
  \textbf{0.013} &
  \textbf{0.0094} \\ \thickhline
\multirow{2}{*}{SSIM$\uparrow$} &
  Original &
  0.9888 &
  0.9044 &
  0.8057 &
  0.6169 &
  0.9701 &
  0.6868 &
  0.9426 &
  0.5672 &
  0.8423 &
  0.7275 \\ \cline{2-12} 
 &
  Sharing Linear &
  0.9633 &
  0.8827 &
  0.8007 &
  \textbf{0.6326} &
  0.9411 &
  \textbf{0.6913} &
  0.9096 &
  \textbf{0.5873} &
  \textbf{0.8512} &
  0.7248 \\ \hline
\end{tabular}
\vskip -0.1in
\caption{Quantitative results of sharing linear converter over multiple datasets, compared with original results. Both the encoder/decoder and the linear layer are shared across each dataset family. We highlight the improvement of the results after sharing the linear converter.}
\label{table:share_linear}
\vskip -0.2in
\end{table*}

\textbf{MAE v.s. Other Self-Supervised Learning Method.}
In a further experiment, we utilized vanilla autoencoders~(i.e., mask ratio equals zero.) and another self-supervised algorithm, Masked FINOLA-B \cite{chen2023image}, with the same architecture as the self-supervised training models. These models were pre-trained on the "Fault Family" dataset and then applied to train and validate a linear converter on CurveFault-A as an illustrative case. The reconstruction and inversion results are shown in Table~\ref{table:autoencoder}. As demonstrated, our framework is adaptable beyond MAE to include other advanced self-supervised learners, as they impose sufficient constraints. However, a simple vanilla autoencoder cannot capture the crucial information that is necessary for both reconstructing and connecting to another domain. If we simply consider both seismic data and velocity maps as pure images and ignore the physical meaning behind them, the vanilla autoencoder would learn too many shortcuts that are only useful to reconstruct the image but lose the essential information reflecting its physical properties. This is because seismic data and velocity maps are not as diverse as natural images. On the other hand, if a model can embed the essential underlying physics information of these two quantities, it will naturally enhance the generalization ability.

\begin{table}[h]
\renewcommand{\arraystretch}{1.3}
\centering
\scriptsize
\setlength{\tabcolsep}{0.8mm}
\begin{tabular}{c|c|c|c|c|c}
\hline
Model &
  MAE$\downarrow$ &
  MSE$\downarrow$ &
  SSIM$\uparrow$ &
  \begin{tabular}[c]{@{}c@{}}Seismic \\ Pre-training \\MAE$\downarrow$\end{tabular} &
  \begin{tabular}[c]{@{}c@{}}Velocity \\ Pre-training\\MAE$\downarrow$\end{tabular} \\ \hline
Masked Autoencoder &
\textbf{0.0277} &
\textbf{0.0061} &
  0.9426 &
  0.1703 &
  0.0410 \\ \hline
Autoencoder &
  0.0614 &
  0.0174 &
  0.8302 &
  \textbf{0.0008} &
  \textbf{0.0005} \\ \hline
Masked FINOLA-B &
  0.0287 &
  0.0062 &
  \textbf{0.9430} &
  0.0083 &
  0.0579 \\ \hline
\end{tabular}
\caption{Comparison between different pre-training strategies on CurveFault-A. In addition to the quantitative results of inversion, the mean absolute reconstruction errors (with masks) of the pre-trained models~(Columns 5 \& 6) are also reported.}
\label{table:autoencoder}
\end{table}

\textbf{Rank of Linear.} We evaluate performances over five different numbers of ranks of the linear converter, varying from 32 to 128. The quantitative results are shown in Table~\ref{table:bottleneck}. Results indicate that increasing the rank makes the model much larger, but the growth of the results is limited.
On the other hand, decreasing the model's rank also does not reduce its capacity a lot but results in a smaller number of parameters. This allows for the balance of performance and computational cost based on specific requirements and available resources, highlighting the flexibility of our model.

\begin{table}[ht]
\scriptsize
\centering
\begin{tabular}{c|c|c|c|c|c}
\thickhline
{ Dataset} & { Dim} & { \#Param} & { MAE$\downarrow$} & { MSE$\downarrow$} & { SSIM$\uparrow$} \\ \thickhline
                               & 128* & 12.3M & 0.0277 & 0.0061 & 0.9426 \\ \cline{2-6} 
                               & 512 & 26.0M & 0.0271 & 0.0058 & 0.9441 \\ \cline{2-6} 
                               & 256 & 16.8M & 0.0274 & 0.0059 & 0.9434 \\ \cline{2-6} 
                               & 64  & 10.0M & 0.0280 & 0.0064 & 0.9392 \\ \cline{2-6} 
\multirow{-5}{*}{CurveFault-A} & 32  & 8.9M  & 0.0304 & 0.0075 & 0.9300 \\ \hline
\end{tabular}
\caption{Quantitative results of different dimensions of bottleneck in the linear converter, and the corresponding number of parameters. As a reference, the number of parameters of InversionNet is 24.4M. (*) indicates the default decoder option.}
\label{table:bottleneck}
\vskip -0.1in
\end{table}

\textbf{Model Complexity and Mask Ratio.} We test the influence of model complexity and self-supervised training hyperparameters (i.e., mask ratio). We test a simpler encoder and smaller latent dimension (namely Auto-Linear-SE), a more complex encoder and larger latent dimension (namely Auto-Linear-LE and Auto-Linear-LD), and corresponding settings for the decoder (namely Auto-Linear-SD and Auto-Linear-LD). For Auto-Linear-SE, we use a one-layer transformer as the encoder and reduce the latent dimension of seismic data to 72.  For Auto-Linear-LE, we use a three-layer transformer as the encoder and increase the latent dimension of seismic data to 264.  For Auto-Linear-SD, we use a one-layer transformer as the decoder and reduce the latent dimension of velocity to 252.  For Auto-Linear-LD, we also use a three-layer transformer as the decoder and increase its width to 768.
We also test a different masking ratio (MR) of 0.5 for the encoder and decoder separately. The results are shown in Table~\ref{table:complexity}.
\begin{table}[h]
\renewcommand{\arraystretch}{1.3}
\centering
\scriptsize
\setlength{\tabcolsep}{1.8mm}
\begin{tabular}{c|c|c|c}
\hline
Model                  & MAE$\downarrow$    & MSE$\downarrow$    & SSIM$\uparrow$   \\ \hline
Auto-Linear                 & 0.0738 & 0.0188 & 0.8057 \\ \hline
Auto-Linear-SE                 & 0.0786 & 0.0206 & 0.7928 \\ \hline
Auto-Linear-LE                 & 0.0685 & 0.0170 & 0.8191 \\ \hline
Auto-Linear-SD                 & 0.0760 & 0.0188 & 0.7880 \\ \hline
Auto-Linear-LD                 & 0.0729 & 0.0194 & 0.8033 \\ \hline
Auto-Linear (encoder MR 0.5)   & 0.0785 & 0.0205 & 0.7941 \\ \hline
Auto-Linear (decoder MR 0.5)   & 0.0744 & 0.0196 & 0.7922 \\ \hline
\end{tabular}
\caption{Quantitative results on CurveVel-A with different hyperparameters.}
\label{table:complexity}
\vskip -0.2in
\end{table}

The results highlight a few key observations: 1) our approach exhibits robustness across various hyperparameter selections; 2) a small model and latent space will relatively influence the model capacity, while a more complex model can achieve even better results. This suggests that our new paradigm has a substantial potential to achieve even better results. Our primary focus is to provide novel insights and introduce a new paradigm. Thus, our choices of hyperparameters are designed to balance results and model complexity, maintaining generality rather than being specifically tailored for a specific dataset. Regarding the masking ratio, for now, we can see that a very high masking ratio (i.e., 0.75) benefits the downstream task, which is consistent with the conclusion in the original MAE paper. Further discussion of the impact of masks in representation learning and whether masking is the best pre-task for image-like data is out of the scope of this paper.
We would love to explore more advanced methods for learning better representations that make the latent-space representations closer to the physical nature in our future work.

\textbf{Patching size.} Our choice for the current patch size ensures a balanced and reasonable patch size and the number of patches, avoiding extremes in either direction. The selected sizes—$(100, 10)$ for seismic and $(10, 10)$ for velocity—uphold broad applicability. We conducted additional experiments on CurveVel-A using two different patch sizes for seismic data: $(250, 10)$ and $(500, 14)$. The results, presented in Table~\ref{table:patch}, indicate that the patch size has a small influence on the final performance. Our model demonstrates robustness across varying patch sizes, reinforcing the logic behind our current selection.

\begin{table}[h]
\renewcommand{\arraystretch}{1.3}
\centering
\scriptsize
\setlength{\tabcolsep}{1.8mm}
\begin{tabular}{c|c|c|c}
\hline
Seismic Patch Size                  & MAE$\downarrow$    & MSE$\downarrow$    & SSIM$\uparrow$   \\ \hline
(100, 10)*         & 0.0738 & 0.0188 & 0.8057 \\ \hline
(250, 10)         & 0.0779 & 0.0203 & 0.7924 \\ \hline
(500, 14)         & 0.0791 & 0.0210 & 0.7891 \\ \hline
\end{tabular}
\caption{Quantitative results on CurveVel-A with different patch sizes. * is the default configuration.}
\label{table:patch}
\end{table}

\textbf{ Model's Sensitivity to the frequency of the wavelet used in seismic data generation.} We conducted an additional test on CurveFault-A. With pretraining the seismic masked autoencoder on seismic data generated by a 15Hz source wavelet, we train the linear converter with input seismic data generated by a 25Hz source wavelet. The results are displayed in Table~\ref{table:freq}. It can be seen that our model, working on the spatial-temporal domain (an image-to-image translation)  rather than the frequency domain, is compatible with varying frequencies. Additionally, our model can also benefit from higher frequencies, which typically offer higher resolution and sensitivity within a certain depth.
\begin{table}[h]
\renewcommand{\arraystretch}{1.3}
\centering
\scriptsize
\setlength{\tabcolsep}{1.8mm}
\begin{tabular}{c|c|c|c}
\hline
Source Frequency                  & MAE$\downarrow$    & MSE$\downarrow$    & SSIM$\uparrow$   \\ \hline
15Hz*         & 0.0277 & 0.0061 & 0.9426 \\ \hline
25Hz         & 0.0181 & 0.0036 & 0.9626 \\ \hline
\end{tabular}
\caption{Quantitative results on CurveFault-A with different source frequencies. * is the default configuration.}
\label{table:freq}
\vskip -0.2in
\end{table}

\subsection{Comparing the latent representations of Auto-Linear and InvLINT.}
To further analyze the relationship between our Auto-Linear and InvLINT, in this part, we compare the latent representations of seismic data and velocity maps obtained by our method to those obtained by InvLINT. First, We conducted experiments on CurveFault-A that use a sine kernel from InvLINT as the encoder and use our pre-trained decoder to construct the inversion network, respectively. The converter is still linear. The results are shown in Table~\ref{table:kernel_enocder}. These results show that using the latent seismic representation from the sine kernel is difficult to regress the latent velocity representation from our method.

\begin{table}[ht]
\scriptsize
\centering
\begin{tabular}{c|c|c|c}
\thickhline
 { Model} & { MAE$\downarrow$} & { MSE$\downarrow$} & { SSIM$\uparrow$} \\ \thickhline
  Auto-Linear & 0.0277 & 0.0061 & 0.9426 \\ \hline
  Sine Kernel Encoder & 0.0426 & 0.0093 & 0.9233 \\ \hline
\end{tabular}
\caption{Comparison between latent representations of seismic data obtained by Auto-Linear and InvLINT on CurveFault-A.}
\label{table:kernel_enocder}
\vskip -0.1in
\end{table}

To further compare the latent representations, we use one latent representation to predict another with linear regression, for seismic data and velocity maps respectively. We report the coefficient of determination ($R^2$ score) in Table~\ref{table:latent_regress}. 
\begin{table}[ht]
\scriptsize
\centering
\begin{tabular}{c|c|c|c}
\thickhline
\multicolumn{1}{l|}{Variable} & Source  & Target  & R\textasciicircum{}2 \\ \thickhline
\multirow{2}{*}{Seismic}       & Auto-Linear  & InvLINT & 0.9869               \\ \cline{2-4} 
                               & InvLINT & Auto-Linear  & 0.6700               \\ \hline
\multirow{2}{*}{Velocity}      & Auto-Linear  & InvLINT & 0.9996               \\ \cline{2-4} 
                               & InvLINT & Auto-Linear  & 0.4871               \\ \hline
\end{tabular}
\caption{Predicting the target latent representations from the source latent representations with linear regression.}
\label{table:latent_regress}
\vskip -0.2in
\end{table}

These show that our latent space with a higher dimension contains more information. As a preliminary comparison, we can roughly conclude that their latent space is a linear subspace of our latent space.

\subsection{Step forward for more complicated seismic models}
Currently, we focus on the acoustic FWI, which is based on one-to-one mapping. For more complicated seismic forward models like the elastic wave equation that includes both P- and S-wave, the scenario shifts to a multiple-input to multiple-output problem.

In light of this, we conducted an additional experiment using our Auto-Linear framework on the elastic FWI dataset, $\mathbf{\mathbb{E}^{FWI}}$ \cite{feng2023efwi}. We trained four Masked Autoencoders (MAEs): two for the different wave types and two for the velocity maps, using the ``Vel Family" dataset. For training the linear converters, we implemented two linear layers to map the embedding of ux and uz into the 128-dimensional bottlenecks, individually. These embeddings were then concatenated and fed through two additional linear layers to produce the latent representations with appropriate dimensions of the respective decoders. The results, shown in Table~\ref{table:efwi}, from the FlatVel-A and CurevVel-A datasets are promising, especially when compared to the ElasticNet benchmark mentioned in the original paper. It is important to note that these are preliminary results, based on our existing one-to-one mapping framework without fine-tuning any hyperparameters. There is still room to explore more effective ways to integrate the four latent spaces, such as how to combine the latent representations from different wave types. However, even at this early stage, our method already achieves a similar level of performance with the supervised benchmarks. This gives us confidence in the potential of the Auto-Linear's applicability for complex seismic models. We will keep exploring these latents' properties across a broader spectrum of tasks in our future work.

\begin{table*}[!h]
\footnotesize
\renewcommand{\arraystretch}{1.3}
\centering
\vspace{0.5em}
\begin{adjustbox}{width=1\textwidth}
\begin{tabular}{c|c|ccc|ccc|ccc} 
\thickhline
\multirow{2}{*}{Dataset}      & \multirow{2}{*}{Model} 
    & \multicolumn{3}{c|}{$\mathrm{V_P}$}
    & \multicolumn{3}{c|}{$\mathrm{V_S}$}
    & \multicolumn{3}{c}{$\mathrm{Pr}$}
    \\ 
\cline{3-11}
    & \multicolumn{1}{c|}{}
    & \multicolumn{1}{c}{MAE$\downarrow$}    & RMSE$\downarrow$    & SSIM$\uparrow$  
    & \multicolumn{1}{c}{MAE$\downarrow$}    & RMSE$\downarrow$    & SSIM$\uparrow$    
    & \multicolumn{1}{c}{MAE$\downarrow$}    & RMSE$\downarrow$    & SSIM$\uparrow$    \\ 
\thickhline
\multirow{2}{*}{$\mathbf{\mathbb{E}^{FVA}}$}
    & Auto-Linear
    & 0.0330    & 0.0672    & 0.9403
    & 0.0241    & 0.0520    & 0.9425    
    & 0.0366    & 0.0752    & 0.7719    \\
    \cline{2-11}
    & ElasticNet
    & 0.0308    & 0.0559    & 0.9615
    & 0.0259    & 0.0500    & 0.9596    
    & 0.0329    & 0.0664    & 0.8455    \\
\thickhline
\multirow{2}{*}{$\mathbf{\mathbb{E}^{CVA}}$}
    & Auto-Linear
    & 0.0781    & 0.1332    & 0.8052
    & 0.0607    & 0.1039    & 0.8081
    & 0.0771    & 0.1280    & 0.4721    \\
    \cline{2-11}
    & ElasticNet   
    & 0.0745    & 0.1345    & 0.8055
    & 0.0600    & 0.1080    & 0.8051
    & 0.0574    & 0.1156    & 0.5766    \\
\thickhline
\end{tabular}
\end{adjustbox}
\caption{Quantitative results of Eelastic FWI. }
\vspace{0.5em}
\label{table:efwi}
\vspace{-1em}
\end{table*}

\subsection{Evaluation on the Recovery of Faint Seep Reflectors}
To further evaluate the performance of our method, we show extra evaluation results of the conventional physics-driven seismic imaging methods.
To evaluate the detection of subtle subsurface reflectors at considerable depths, we apply the reverse time migration (RTM) technique. RTM is a computational approach utilized in seismic imaging, enabling the creation of high-resolution visualizations of subsurface structures beneath the Earth's surface. Besides, we perform a zero-offset least square reverse-time migration (LSRTM) of the predicted velocity maps under the Born approximation scenario in order to show a evaluation of the recovery of deep reflectors. The LSRTM is calculated for one predicted sample from each dataset used in the paper, with 20 optimization iterations. 

\begin{figure}[ht]
\vskip 0.2in
\begin{center}
\centerline{\includegraphics[width=.7\columnwidth]{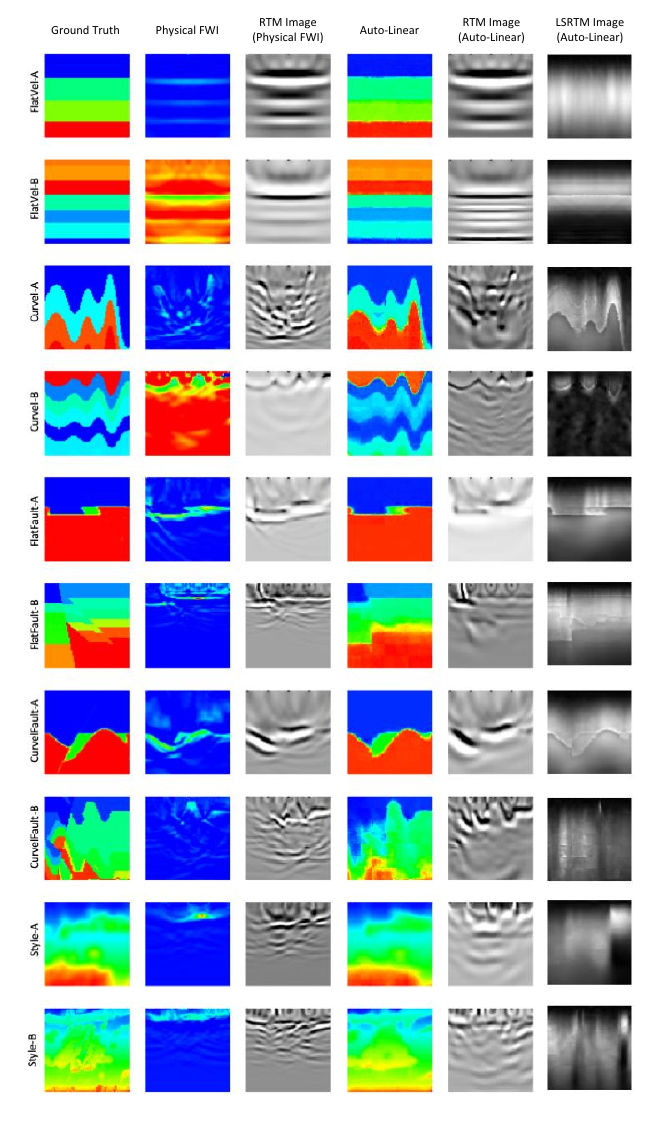}}
\caption{Ground truth, velocity maps and RTM images obtained with physics-driven FWI versus velocity maps and RTM images obtained with Auto-Linear} 
\label{Auto-Linear_RTM}
\end{center}
\vskip -0.4in
\end{figure}

For comparison purposes, we also conduct physics-driven FWI \cite{schuster2017seismic} alongside its corresponding RTM. As Auto-Linear obviates the need for an initial model and avoids data preprocessing in this study, to ensure an equitable comparison, the physical FWI is executed with a uniform background and data generated using a 15 Hz Ricker wavelet without applying any bandpass filtering. The outcomes are presented in Figure~\ref{Auto-Linear_RTM}. It is evident that Auto-Linear outperforms Physical FWI in restoring the velocity maps. However, due to the survey's restricted aperture, certain deep reflectors exhibit suboptimal recovery. Nonetheless, the RTM images produced through Auto-Linear still offer improved accuracy and finer details regarding the reflector positions compared to those obtained via physical FWI.

\subsection{Visualizations}

\begin{figure}[!h]
\begin{center}
\centerline{\includegraphics[width=.6\columnwidth]{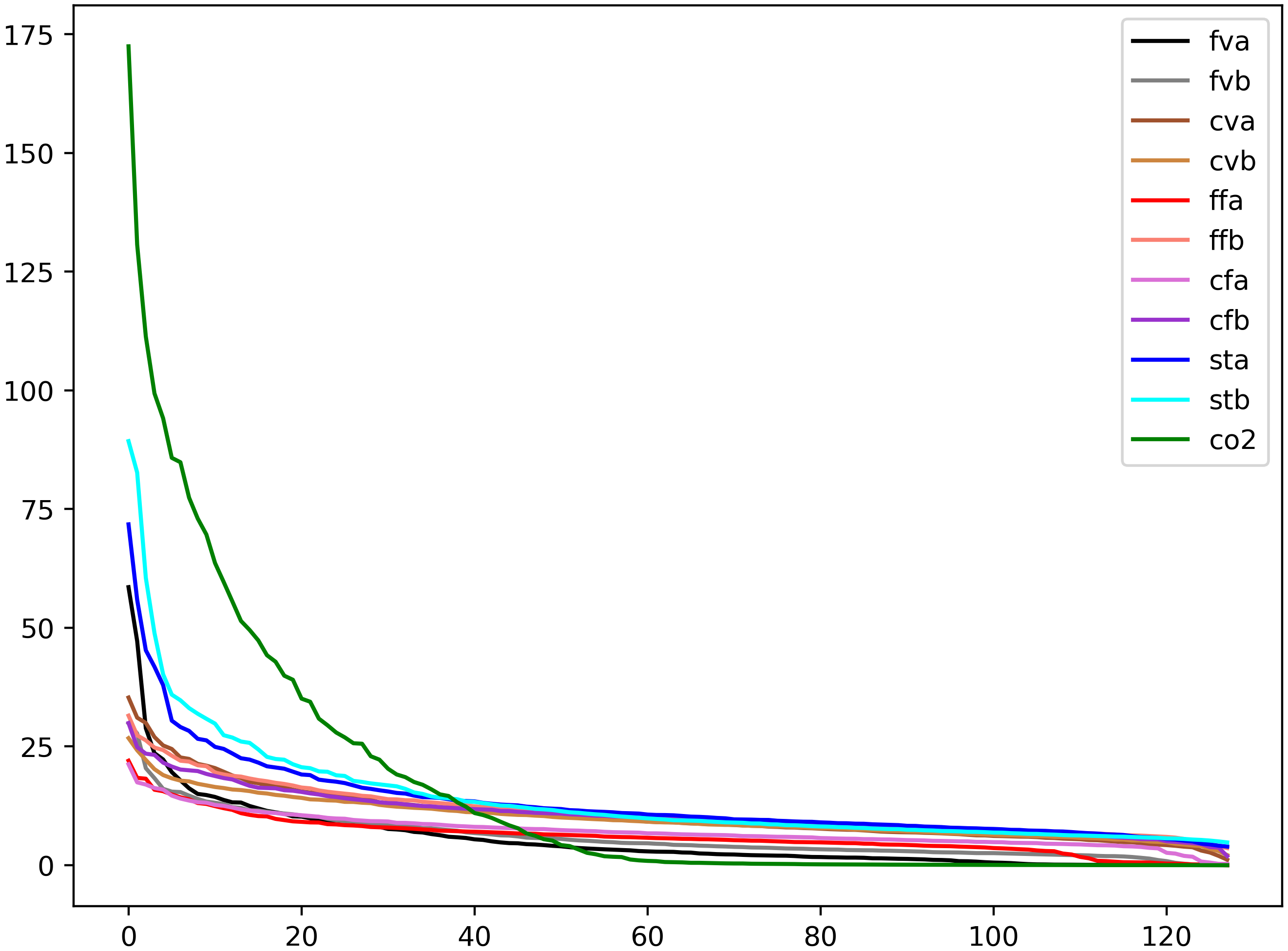}}
\caption{Original Results of Singular Value Decomposition on different datasets.} 
\label{svd2}
\end{center}
\end{figure}

\begin{figure}[!h]
\begin{center}
\centerline{\includegraphics[height=.98\textheight]{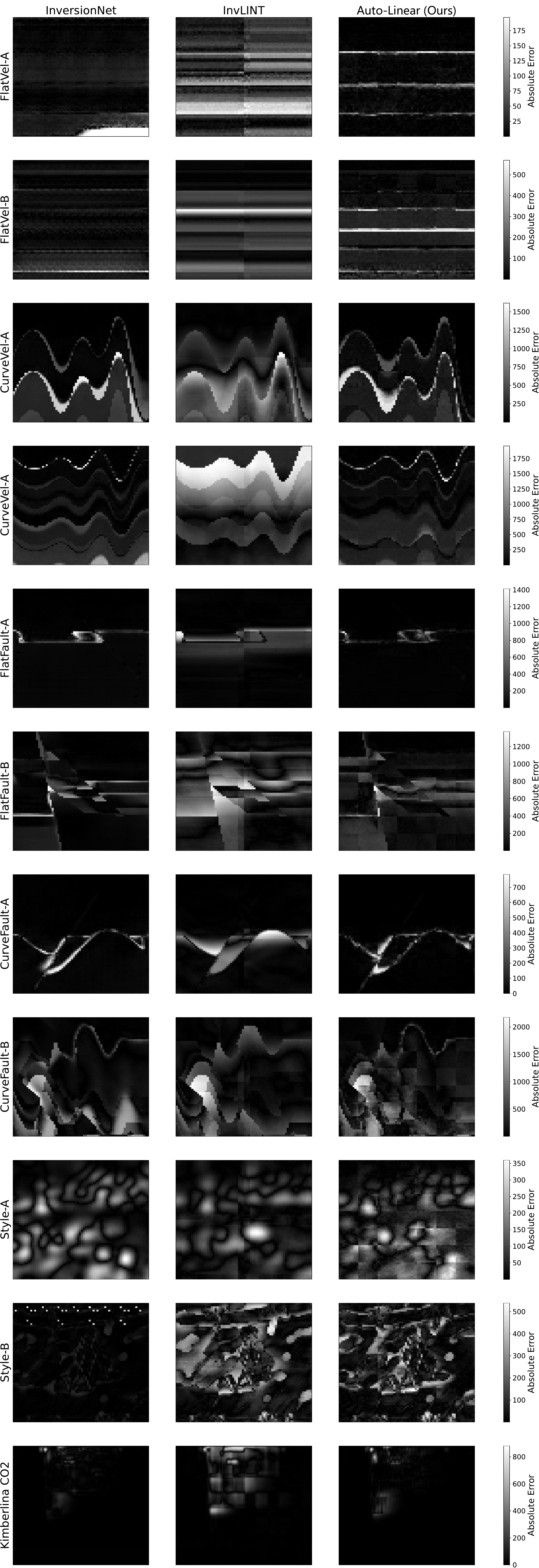}}
\vskip -0.1in
\caption{Illustration of absolute error map on OpenFWI, compared, Auto-Linear, InversionNet~\cite{wu2019inversionnet} and InvLINT~\cite{feng2022intriguing} to the ground truth.} 
\label{main_result1_2}
\end{center}
\vskip -0.5in
\end{figure}

\begin{figure}[!h]
\begin{center}
\centerline{\includegraphics[height=.98\textheight]{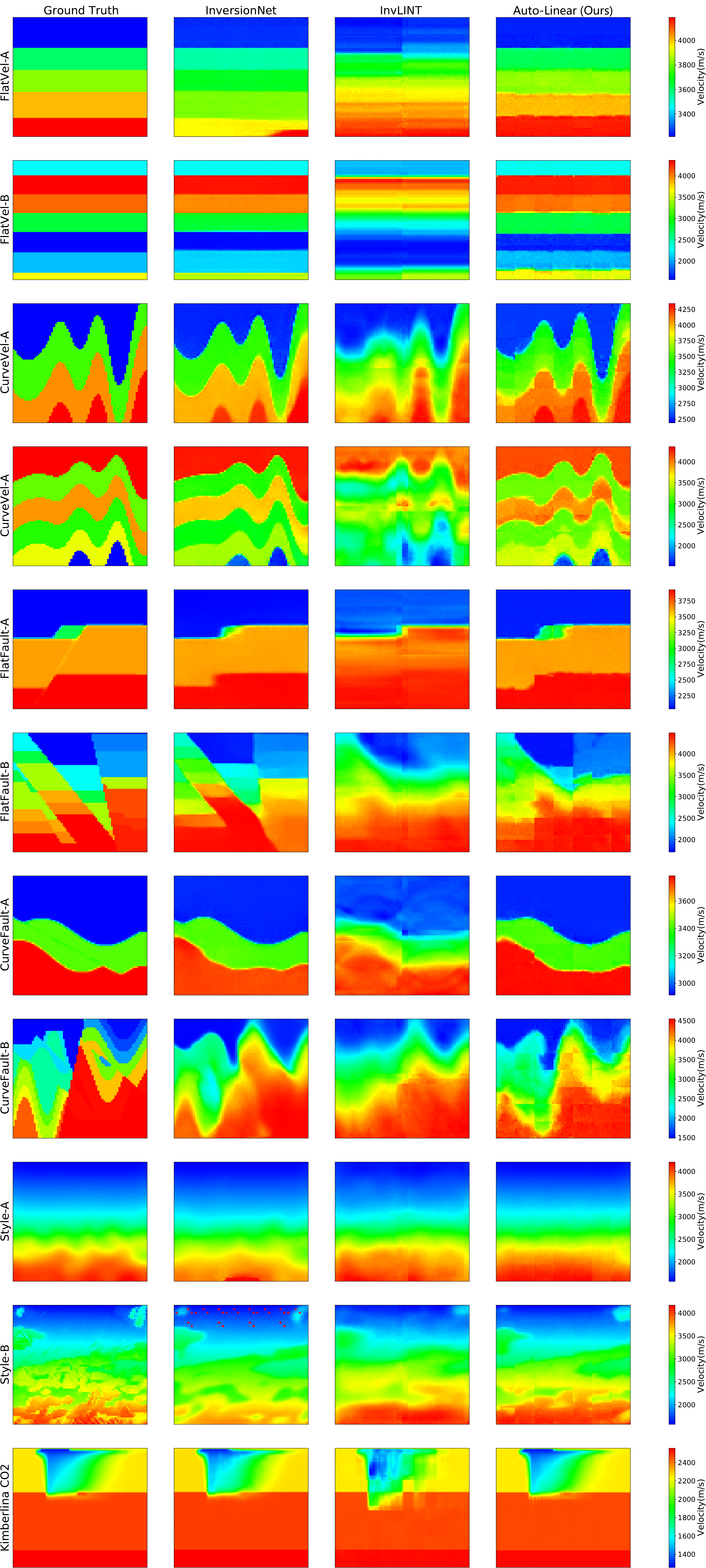}}
\vskip -0.1in
\caption{Illustration of results evaluated on OpenFWI, compared with InversionNet~\cite{wu2019inversionnet} and InvLINT~\cite{feng2022intriguing}.} 
\label{main_result2}
\end{center}
\vskip -0.5in
\end{figure}

\begin{figure}[!h]
\begin{center}
\centerline{\includegraphics[height=.98\textheight]{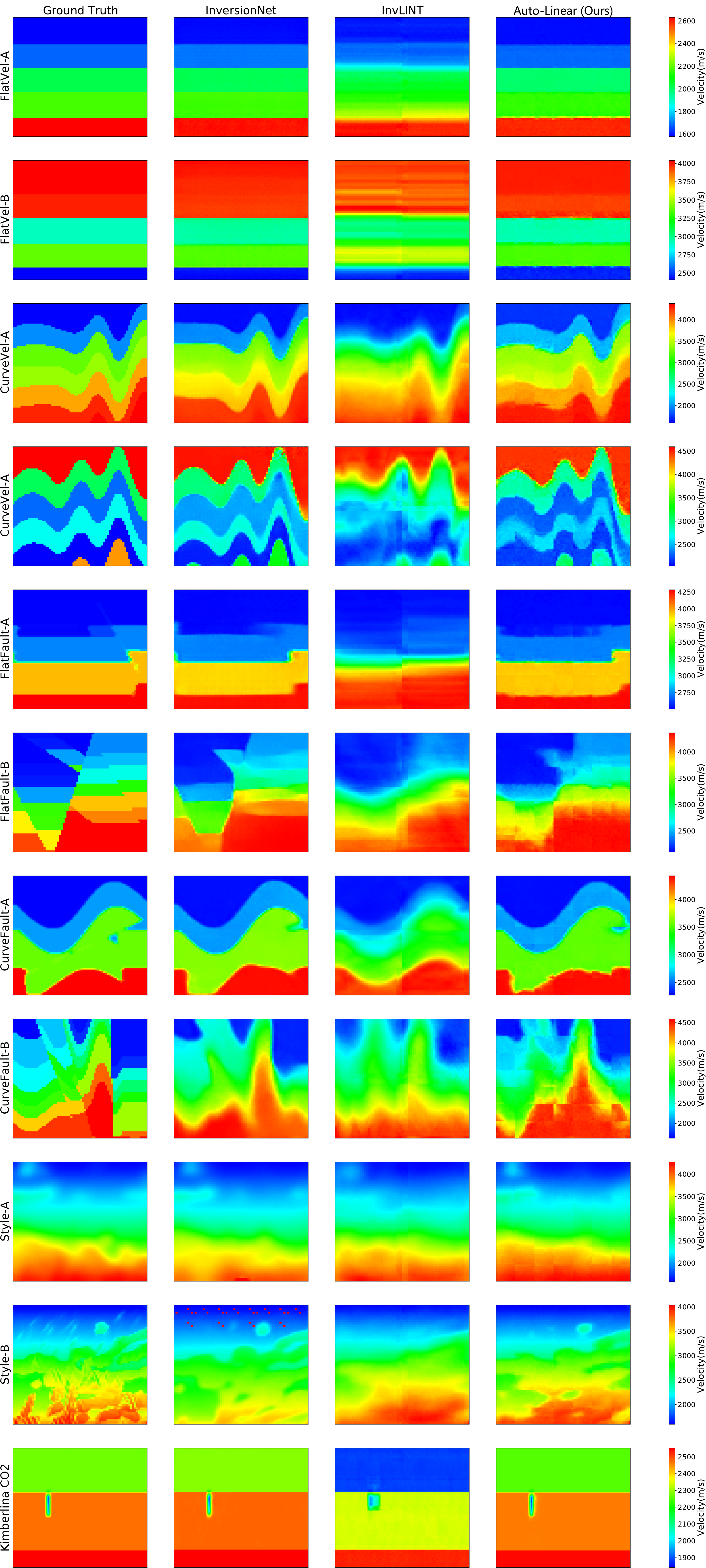}}
\vskip -0.1in
\caption{Illustration of results evaluated on OpenFWI, compared with InversionNet~\cite{wu2019inversionnet} and InvLINT~\cite{feng2022intriguing}.} 
\label{main_result3}
\end{center}
\vskip -0.5in
\end{figure}

\end{document}